\documentclass[aps,pre,twocolumn,superscriptaddress,amsmath,amssymb]{revtex4-1}
\pdfoutput=1
\usepackage{times}
\usepackage{array}
\usepackage{multirow}
\usepackage{graphicx}
\usepackage{bm}
\usepackage{adjustbox}
\usepackage{xcolor,colortbl}
\usepackage{tikz,colortbl}
\usetikzlibrary{calc}
\usepackage{zref-savepos}
\usepackage{multirow}

\definecolor{bblue}{RGB}{123,124,203}
\definecolor{lbblue}{RGB}{184,182,245}
\definecolor{ggreen}{RGB}{143,209,142}
\definecolor{lggreen}{RGB}{197,231,196}
\definecolor{llggreen}{RGB}{226,243,226}
\definecolor{lllggreen}{RGB}{237,248,234}
\definecolor{lgrey}{RGB}{230,230,230}
\definecolor{dgrey}{RGB}{200,200,200}

\newlength{\Oldarrayrulewidth}

\newcommand{\tikzmark}[1]{\tikz[remember picture,overlay, baseline=-1.5ex]\node (#1){};}
\newcommand{\connect}[3][3mm]{\tikz[remember picture,overlay]\draw[shorten <=-#1, shorten >=-#1](#2)--(#3);}

\newcounter{NoTableEntry}
\renewcommand*{\theNoTableEntry}{NTE-\the\value{NoTableEntry}}

\newcommand*{\notableentry}{%
  \multicolumn{1}{@{}c@{}|}{%
    \stepcounter{NoTableEntry}%
    \vadjust pre{\zsavepos{\theNoTableEntry t}}
    \vadjust{\zsavepos{\theNoTableEntry b}}
    \zsavepos{\theNoTableEntry l}
    \hspace{0pt plus 1filll}%
    \zsavepos{\theNoTableEntry r}
    \tikz[overlay]{%
      \draw[black]
        let
          \n{llx}={\zposx{\theNoTableEntry l}sp-\zposx{\theNoTableEntry r}sp},
          \n{urx}={0},
          \n{lly}={\zposy{\theNoTableEntry b}sp-\zposy{\theNoTableEntry r}sp},
          \n{ury}={\zposy{\theNoTableEntry t}sp-\zposy{\theNoTableEntry r}sp}
        in
        (\n{llx}, \n{lly}) -- (\n{urx}, \n{ury})
        (\n{llx}, \n{ury}) -- (\n{urx}, \n{lly})
      ;
    }%
  }%
}

\begin{document}
\title{Partial Synchronization and Partial Amplitude Death in Mesoscale Network Motifs}

\author{Winnie Poel}
\author{Anna Zakharova}
\author{ Eckehard Sch{\"o}ll}
\email{ schoell@physik.tu-berlin.de}
\affiliation{Institut f{\"u}r Theoretische Physik, Technische
Universit{\"a}t Berlin, Hardenbergstr. 36, 10623 Berlin, Germany}

\begin{abstract}
We study the interplay between network topology and complex space-time patterns and introduce a concept to analytically predict complex patterns in networks of Stuart--Landau oscillators with linear symmetric and instantaneous coupling based solely on the network topology.
These patterns consist of partial amplitude death and partial synchronization and are found to exist in large variety for all undirected networks of up to 5 nodes. The underlying concept is proved to be robust with respect to frequency mismatch and can also be extended to larger networks. In addition it directly links the stability of complete in-phase synchronization to only a small subset of topological eigenvalues of a network.
\end{abstract}
\maketitle

\section{Introduction}
In recent years the study of networks has increasingly improved the understanding of complex systems across scientific fields. Interpreting a complex system as a network whose nodes are nonlinear oscillators interacting via the edges of the network graph, one can discover an abundance of collective phenomena that have also been widely observed in experiments. Among others, global synchronization and oscillation quenching mechanism, i.e., amplitude death (AD) and oscillation death (OD), have been intensely studied both theoretically and experimentally \cite{PIK03,SAX12,KOS13}. Amplitude death is associated with the stabilization of an already existing trivial steady state while oscillation death is characterized by a newly born inhomogeneous steady state. Applications of synchronization range from neuronal and genetic networks to coupled electrical and laser networks, coupled mechanical systems or even networks in social sciences \cite{PIK03,SUY08,SIN99a,VAR01,WIN90,NED00,WIT12} while oscillation quenching has been observed across many man-made and natural systems ranging from lasers and electronic circuits \cite{KIM05,HEI10}, chemical and biological networks including neurons \cite{CRO98,DOL88,CUR10} to climate systems \cite{GAL01}. Applications of AD are mainly in controlling physical and chemical systems (e.g., coupled lasers \cite{KUM08}) and suppressing neuronal oscillations \cite{ERM90a,CAK14}, while OD has been suggested as a mechanism to generate heterogeneity in homogeneous systems (e.g., stem cell differentiation \cite{SUZ11} in morphogenesis). Synchronized behavior is in some cases considered essential for the proper functioning of the network (e.g., power distribution networks \cite{MOT13a} or coherent circadian output in mammals \cite{MOO97, YAM03a}) while in others it is undesired and harmful (e.g., in several pathological neuronal states such as Parkinson's disease and essential tremor \cite{TAS98,MIL03a}). Thus the control of synchronization has also widely been studied \cite{CHO09,SCH12,SCH13,LEH14}.

The extensions of the studies of these global phenomena towards cluster and group synchronization \cite{SOR07,DAH12,WIL13,PEC14,TAY11a,TIN12,NKO13}, partial amplitude and oscillation death \cite{ATA02a,KOS13,ATA02a,RUB02} and chimera states \cite{KUR02a,ABR04,OME13,PAN14} are first steps towards a better understanding of the emergence of local mesoscale structures on networks and their influence on the global dynamics \cite{DO12}.

 The identical temporal behavior of parts of a network is common to all of these phenomena and might be a universal concept in coupled systems that one can try to understand independently of the dynamical phenomenon itself on the basis of the network topology. Efforts have recently been made to understand the spectral properties and symmetries of complex networks \cite{MAC09a,MAC08,XIA08}, and from some of the results dynamical implications have been inferred \cite{PEC14,DO12}. Ground breaking work connecting the topological properties of the network with dynamical features was the introduction of the master stability function for the completely synchronous state \cite{PEC98} and its extension to time-delayed coupling \cite{KIN09,CHO09,FLU10b,SOR13} and group and cluster synchronization \cite{SOR07,DAH12,PEC14}.

The interplay of local dynamics and the network topology and their role in the formation of dynamic patterns is nevertheless only poorly understood in general, and a systematic and large-scale comparison of the different dynamical patterns which a certain network topology can display has not been made. 
Two of the issues complicating this task are the complex local dynamics each oscillator already exhibits on its own and the complexity and sheer size of real world networks.

This paper addresses this question of a possible connection between the network topology and the dynamical patterns it can support. The model system used for this purpose should be as simple as possible so that general basic mechanisms may become apparent. Therefore a symmetric instantaneous coupling and identical elements with simple generic local dynamics will be considered. The Stuart-Landau oscillator is such a generic model of a nonlinear oscillator close to a Hopf bifurcation. It has been widely studied, and shows the above phenomena, i.e., synchronization, amplitude and oscillation death, and chimera states \cite{MAT90,MIR90,ATA03,ZAK13a,ZAK14,SET14}. It is used for the local dynamics of the model in this paper.

The size of the network is the other factor making it difficult to understand general mechanisms clearly. The study of network motifs has suggested that these small networks may have dynamical implications or functions on their own \cite{MIL02,VAZ04,VAL05,DO12}. Therefore, this paper uses the whole variety of undirected motifs with up to five nodes as a reference system for the occurrence of dynamical patterns. Having such an abundance of examples helps to distinguish properties that belong to one specific topology only from general mechanisms. 

The paper is structured as follows. Sect. \ref{sec_model} introduces the model we use. Sect. \ref{sec_eigensols} presents the analytical approach developed in our work that can be used to deduce dynamical patterns from a network's topology. The stability of these patterns is analyzed in Sect. \ref{sec_stabana} before their robustness with respect to mismatch in the oscillators' frequencies is studied in Sect. \ref{sec_het}. Finally, Sect. \ref{sec_tcn} points out how to extend our concept to larger networks.

\section{Model} \label{sec_model}
We use the Stuart-Landau oscillator, a generic model of a system close to a Hopf bifurcation. The dynamics of the $i$-th oscillator, $z_i\in\mathbb{C},~i\in\{1,\dots,N\} $ is given by
\begin{equation}
\dot{z_i}=f(z_i)+\sigma e^{i\beta}\sum_{j=1}^N A_{ij}z_j
\label{eq:model}
\end{equation}
\begin{equation}
f(z)=(\lambda+i\omega-|z|^2)z
\label{eq:f}
\end{equation}
where $\omega\in\mathbb{R}$ is the oscillator frequency, and $\lambda\in\mathbb{R}$ is the bifurcation parameter. For $\lambda>0$ there exists a limit cycle of radius $\sqrt{\lambda}$ that is born in a supercritical Hopf bifurcation at $\lambda=0$ for a single Stuart--Landau oscillator. The real parameters $\sigma$ and $\beta$ are the coupling strength and coupling phase, respectively.

\begin{figure}[h]
a)\adjustbox{valign=t}{\includegraphics[scale=0.2]{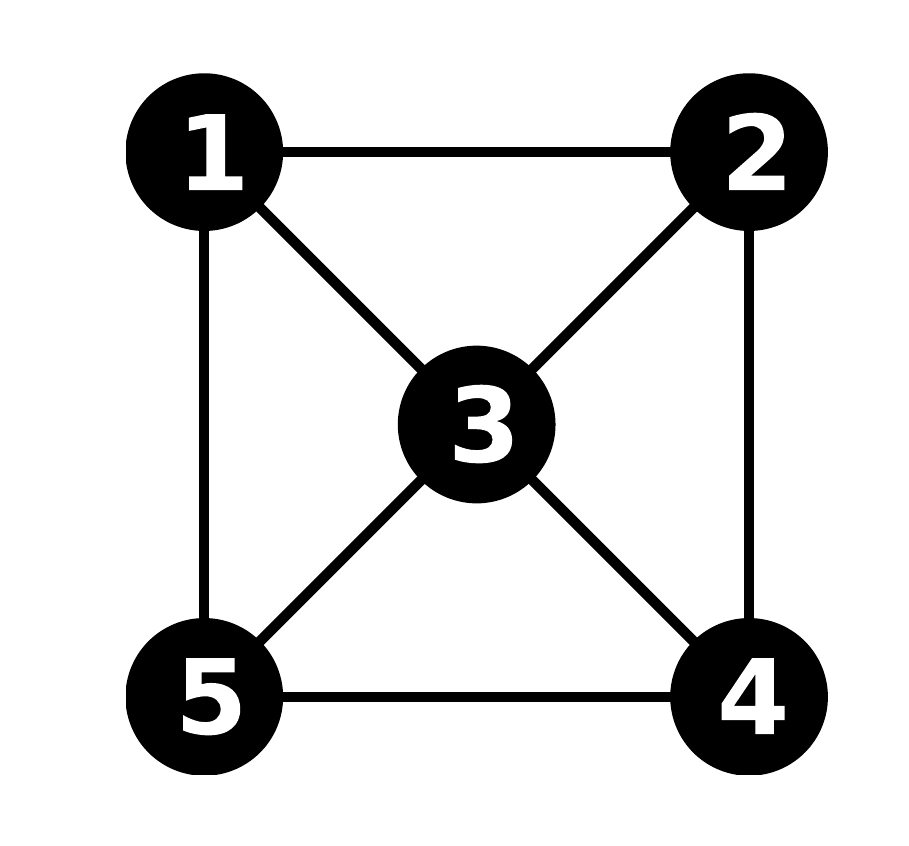}}\hspace{0.4cm}
b)\hspace{0.2cm}\adjustbox{valign=t}{$\bf{A}=\left(
\begin{array}{ccccc}
 0 & \frac{1}{3} & \frac{1}{3} & 0 & \frac{1}{3} \\
 \frac{1}{3} & 0 & \frac{1}{3} & \frac{1}{3} & 0 \\
 \frac{1}{4} & \frac{1}{4} & 0 & \frac{1}{4} & \frac{1}{4} \\
 0 & \frac{1}{3} & \frac{1}{3} & 0 & \frac{1}{3} \\
 \frac{1}{3} & 0 & \frac{1}{3} & \frac{1}{3} & 0 \\
\end{array}
\right)$}
\caption{\textbf{a)} Motif of 5 coupled oscillators given by the normalized adjacency matrix, A, shown in \textbf{b)}. It is used in this paper for illustrating general results that can be applied to all undirected motifs.}
\label{fig:motif519}
\end{figure}

The oscillators are connected via a linear bidirectional instantaneous coupling using an adjacency matrix, $A$, with unity row sum,
\begin{equation}
\sum_{j=1}^NA_{ij}=1~,
\end{equation}
to ensure that complete in-phase synchronization can be reached.
This matrix $A$ has only real eigenvalues $-1\leq\eta\leq 1$ (proof see appendix A).
An example of this normalized adjacency matrix can be found in Fig. \ref{fig:motif519}b) for a five-node motif.
In this paper we focus on the motif depicted in Fig. \ref{fig:motif519}a) as an illustrative example, but similar results have been obtained for all undirected network motifs with up to $N=5$ nodes in our work \footnote{See Supplemental Material at [] for details on the other motifs.}.

\section{Eigensolutions -- An Analytic Concept} \label{sec_eigensols}
This section introduces an analytic solution for the temporal behavior of the oscillators $z_i(t)$ that are described by the coupled ordinary differential equations (ODEs) in Eq. \eqref{eq:model}. It is obtained by using the following ansatz. An eigenvector $\mathbf{v}$ of the normalized adjacency matrix $A$ with eigenvalue $\eta$
\begin{equation}
A\mathbf{v}=\eta\mathbf{v}~.
\label{eq:aev}
\end{equation}
and eigenvector components 
\begin{equation}
v_i\in\{-1,0,+1\}
\end{equation} is used to make the ansatz
\begin{equation}
z_i(t)=v_iz_\eta(t)~.
\end{equation}
The function $z_\eta(t)$ is to be determined.
The following properties of the Stuart-Landau dynamics
\begin{equation}
\begin{aligned}
 f(0)&=0\\
 f(-z_i)&=-f(z_i)
\end{aligned}
 \label{eq:fspecial}
\end{equation}
 are combined in
\begin{equation}
f(z_\eta v_i)=v_if(z_\eta)
\label{eq:fobs}
\end{equation}
and used to decouple the system of ODEs in Eq. \eqref{eq:model} yielding
\begin{equation}
 \dot{z}_\eta v_i=\big(f(z_\eta)+\sigma e^{i\beta}z_\eta\eta\big) v_i~.
 \label{eq:z_eta}
\end{equation}
If $v_i=0$ then Eq. (\ref{eq:z_eta}) is trivially true independently of $z_\eta$. For $v_i\neq0$ dividing it by $v_i$ yields
\begin{equation}
 \dot{z}_\eta=f(z_\eta)+\eta\sigma e^{i\beta}z_\eta~.
 \label{eq:egal}
\end{equation}
We introduce new parameters
\begin{equation}
\begin{aligned}
\tilde{\lambda}&=\lambda+\eta\sigma\cos\beta\\
 \tilde{\omega}&=\omega+\eta\sigma\sin\beta
 \end{aligned}
\label{eq:tildelamomdef}
\end{equation}
and using that $\eta\in\mathbb{R}$ one can write Eq. (\ref{eq:egal}) as
\begin{equation}
\begin{aligned}
\dot{z}_\eta&=(\tilde{\lambda}+i\tilde{\omega}-|z_\eta|^2)z_\eta~.
\end{aligned}
\end{equation}
This has the form of the local dynamics of a single Stuart-Landau oscillator given in Eq. (\ref{eq:f}) and is solved by $z_\eta\equiv z_\eta^*=0$ or the limit cycle
\begin{align}
 z_\eta\equiv z_\eta^{LC}=\sqrt{\tilde{\lambda}}e^{i\tilde{\omega}t}~.
 \label{eq:zeta}
\end{align}
In summary the system has, in addition to the trivial steady state, $z_\eta^*=0$, a new periodic attractor given by
\begin{align}
z_i(t)=v_i\;z_\eta^{LC}(t)~.
\label{eq:solution2}
\end{align} 
if the motif's topology is such that the adjacency matrix has an eigenvector consisting only of components $v_i\in\{-1,0,1\}$ with eigenvalue $\eta$.
All considered topologies (undirected motifs up to $N=5$) have at least one eigenvector of the special type needed for the eigensolution. Most motifs exhibit more than one such eigenvector, and therefore more than one eigensolution can be found for them.

In what follows, we explain how the interplay between topology and complex space-time patterns becomes accessible by the concept of an eigensolution.

\begin{figure}[h]
\adjustbox{valign=c}{\includegraphics[scale=0.3]{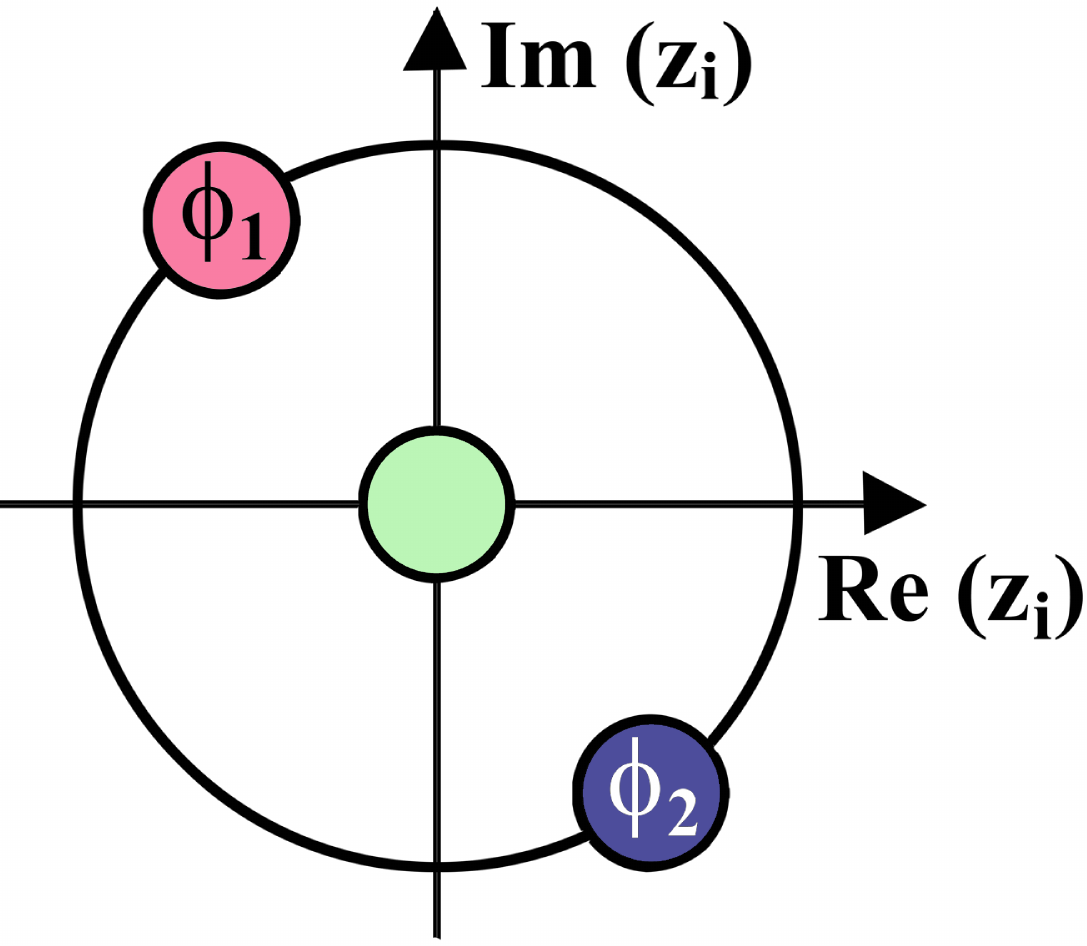}}\hspace{0.5cm}
\begin{minipage}{4cm}
$\begin{aligned}
\adjustbox{valign=c}{\includegraphics[scale=0.4]{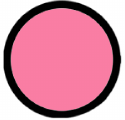}}\hspace{0.2cm}r&=\sqrt{\tilde{\lambda}},\phi=\phi_1\\
\adjustbox{valign=c}{\includegraphics[scale=0.4]{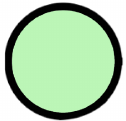}}\hspace{0.2cm}r&=0\\
\adjustbox{valign=c}{\includegraphics[scale=0.4]{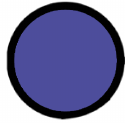}}\hspace{0.2cm}r&=\sqrt{\tilde{\lambda}},\phi=\phi_2=\phi_1+\pi
\end{aligned}$
\end{minipage}
\caption{(color online) Sketch of the different oscillator groups occurring in an eigensolution. The red (dark grey) and blue (black) group are on the limit cycle given by Eq. \eqref{eq:zeta} with phase difference $|\phi_1-\phi_2|=\pi$. Oscillators of the green (light gray) group are in the steady state at $z_i=0$.}
\label{fig:groups}
\end{figure}

According to the three different values of $v_i$, the oscillators $z_i$ can be partitioned into three groups with different temporal behavior as sketched in Fig. \ref{fig:groups}. All oscillators of one group behave identically, i.e., they are fully synchronized in amplitude and phase. The oscillators for which $v_i=0$ holds (green/light gray group), are in a steady state $z_i=0$. Oscillators with $v_i=\pm1$ (red/dark gray and blue/black group) are on the limit cycle $z_\eta$ given by Eq. \eqref{eq:zeta} with a fixed phase difference of $\phi_1-\phi_2=\pi$ between the two groups. Thus eigensolutions are spatial patterns of amplitude death (green/light gray group) and partial in-phase and anti-phase synchronization (blue/black and red/dark gray). We call two oscillators, $z_i=r_ie^{i\phi_i}$ and $z_j=r_je^{i\phi_j}$ with $r_i=r_j$, in-phase synchronized if $|\phi_i-\phi_j|=0$ and anti-phase synchronized if $|\phi_i-\phi_j|=\pi$.

\begin{figure}[h]
a)\hspace{0.05cm}\adjustbox{valign=t}{\includegraphics[scale=0.17]{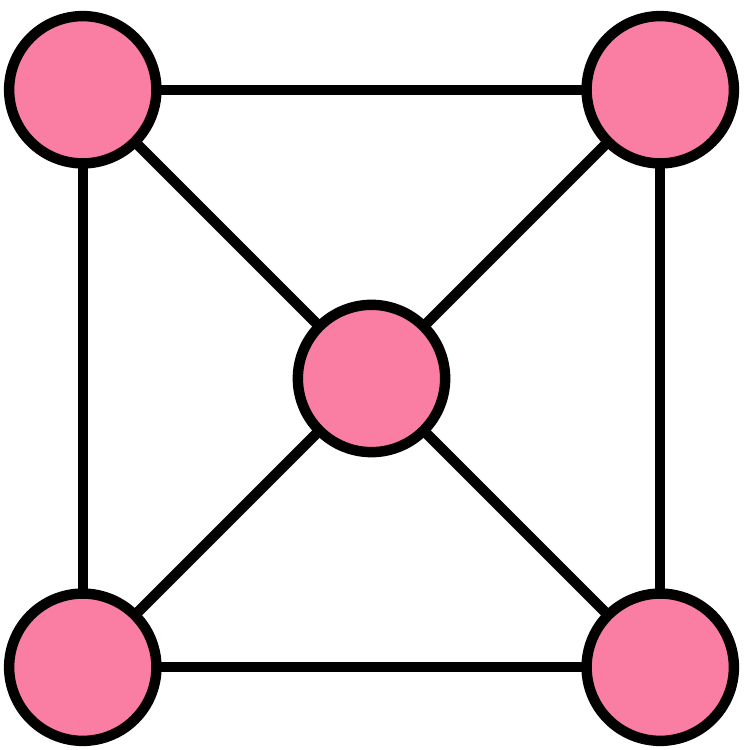}}\hspace{0.09cm}
b)\hspace{0.05cm}\adjustbox{valign=t}{\includegraphics[scale=0.17]{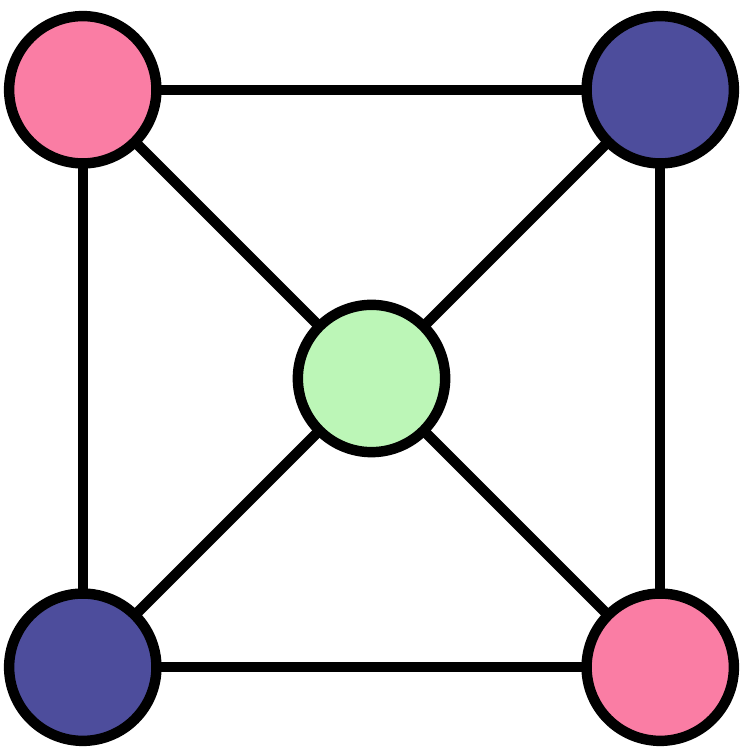}}\hspace{0.09cm}
c)\hspace{0.05cm}\adjustbox{valign=t}{\includegraphics[scale=0.17]{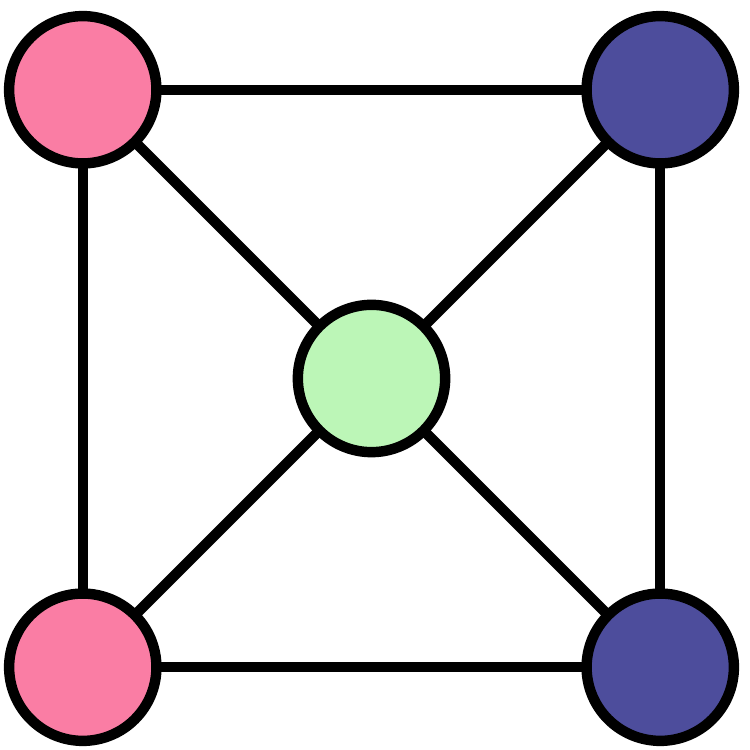}}\hspace{0.09cm}
d)\hspace{0.05cm}\adjustbox{valign=t}{\includegraphics[scale=0.17]{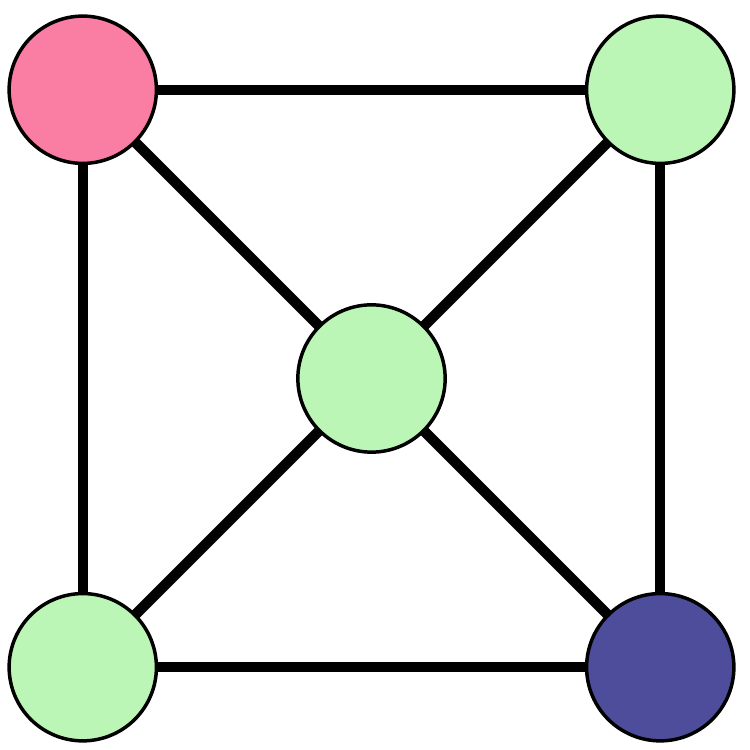}}\hspace{0.09cm}
e)\hspace{0.05cm}\adjustbox{valign=t}{\includegraphics[scale=0.17]{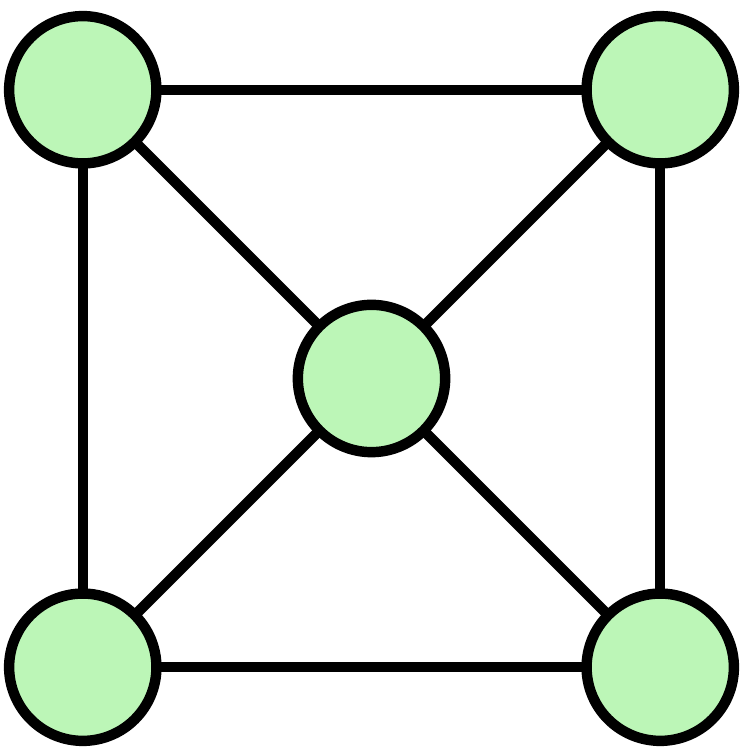}}\hspace{0.09cm}\\\vspace{0.1cm}
\hspace{0.3cm}CIS\hspace{0.88cm}PAD+PS 1\hspace{0.4cm}PAD+PS 2\hspace{0.4cm}PAD+PS 3\hspace{0.7cm}CAD
\caption{(color online) All eigensolutions for one motif as defined by the eigenvectors in Eq. \eqref{eq:eigensystem519}: a) $\mathbf{v^1}$, b) $\mathbf{v^2}$, c) $\mathbf{v^3}+\mathbf{v^4}$, d) $\mathbf{v^4}$. The eigensolutions are patterns of partial amplitude death (green or gray) and partial synchronization (red or light gray, and blue or dark gray). For more details on the color code refer to legend in Fig. \ref{fig:groups}. Additionally, the steady state at $z_i=0~\forall~i\in\{1,\dots,N\}$ is shown in e). These patterns will further be referred to by the acronyms CIS (complete in-phase synchronization), PAD (partial amplitude death), PS (partial synchronization), CAD (complete amplitude death).}
\label{fig:expatterns}
\end{figure}

The spatial distribution of amplitude death and partial synchronization is governed solely by the eigenvector defining the eigensolution and is therefore purely determined by the topology. The motif shown in Fig. \ref{fig:motif519} has the following eigenvectors $\mathbf{v^k}$ and corresponding topological eigenvalues $\eta^k$

\begin{equation}
\begin{aligned}
\mathbf{v^1}&=(1,1,1,1,1)^T &\eta^1&=1\\
\mathbf{v^2}&=(1,-1,0,1,-1)^T &\eta^2&=-2/3\\
\mathbf{v^3}&=(0,1,0,0,-1)^T &\eta^3&=0\\
\mathbf{v^4}&=(1,0,0,-1,0)^T &\eta^4&=0\\
\mathbf{v^5}&=(1,1,-3,1,1)^T &\eta^5&=-1/3~.
\end{aligned}
\label{eq:eigensystem519}
\end{equation}
Of these eigenvectors all but $\mathbf{v^5}$ can be used to obtain an eigensolution. They yield the patterns shown in Fig. \ref{fig:expatterns} a) to d). The pattern shown in c) is a linear combination of the eigenvector $\mathbf{v^3}$ and $\mathbf{v^3}$, which are equivalent if one renumbers the oscillators, and observes that they have the same eigenvalue $\eta=0$. In addition, the system can be in the trivial steady state, shown in part e) of Fig. \ref{fig:expatterns}.

\begin{figure}[h]
\includegraphics[scale=0.56]{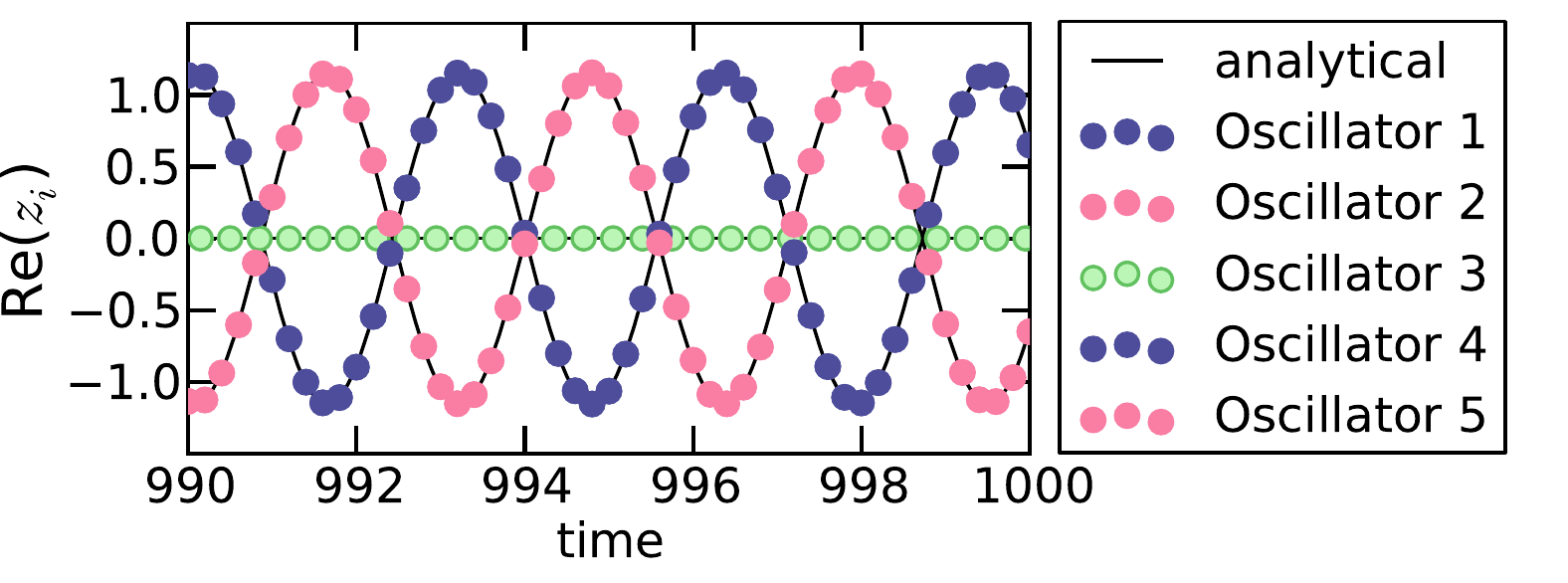}
\caption{(color online) Time series for the PAD+PS 1 pattern shown in Fig. \ref{fig:expatterns}b). The black line gives the analytical eigensolution while the colored dots are the result of numerically integrating the coupled ODEs from Eq. \eqref{eq:model} from random initial conditions up to $t=1000$ time steps. Oscillator 3 is in the steady state at $z_3=0$ while the others are on the limit cycle $z_\eta^{LC}$ given by Eq. \eqref{eq:zeta}. Oscillator 1 and 4 have a fixed phase difference of $\Delta\phi=\pi$ to oscillator 2 and 5. Parameters: $\beta=0, \sigma=-5, \lambda=-2, \omega=2$.}
\label{fig:timeseries}
\end{figure}

A time series of the pattern PAD+PS 1 (Fig. \ref{fig:expatterns}c)) for each oscillator is shown in Fig. \ref{fig:timeseries}. The separation of the oscillators into three groups is clearly visible. Oscillators of the blue (black) and the red (dark gray) group oscillate in anti-phase on the limit cycle while the green (light gray) one is in the steady state at $z_i=0$.
As can be seen in Eqs. \eqref{eq:tildelamomdef} and \eqref{eq:zeta}, the limit cycle of the coupled system has a direct dependence on the topological eigenvalue $\eta$ via its radius $r^{LC}=\sqrt{\tilde{\lambda}}$ and phase $\phi^{LC}=\tilde{\omega}t$. In comparison to the single Stuart-Landau oscillator's limit cycle, $z^{SL}=\sqrt{\lambda}e^{i\omega t}$, the coupled system's radius is altered by the additional term $\eta\sigma\cos\beta$ and its frequency is changed by $\eta\sigma\sin\beta$. 

Note that the example in Fig. \ref{fig:timeseries} is for negative $\lambda$ where the single oscillator is in a stable steady state. The limit cycle of the coupled system is shifted by the additional term in $\tilde{\lambda}$, and therefore the coupled system can exhibit oscillations even though the uncoupled system is in a stable steady state. In general the existence of an eigensolution is determined by $\tilde{\lambda}\geq 0$ and thus
\begin{equation}
\lambda+\eta\sigma\cos\beta\geq 0~.
\label{eq:existence}
\end{equation}
This inequality defines a region in the parameter space of $\lambda$, $\sigma$ and $\beta$ in which the eigensolution corresponding to $\eta$ exists. Since one motif may have several eigensolutions with different topological eigenvalues, it crucially depends on the point in parameter space which spatio-temporal patterns a given motif can exhibit. 

In conclusion, this section has shown that eigensolutions are an analytical method to infer complex space-time patterns from topology. They are built from three groups of oscillators, one being amplitude death and the other two oscillating in anti-phase. The spatial ordering of these groups depends on a topological eigenvector while temporal dynamics and existence in parameter space depend on the corresponding topological eigenvalue.

\section{Linear Stability Analysis} \label{sec_stabana}
In this section, we study the stability of the patterns constructed in Sect. \ref{sec_eigensols}.
We start with patterns without partial amplitude death (PAD) which can be treated in polar coordinates in contrast to the case including PAD where the dead oscillator's phase is undefined. These will be treated separately in Sect. \ref{subsec_cartesian}. Finally Sect. \ref{subsec_tss} analyzes the stability of the trivial steady state, and our results are illustrated for an example in Sect. \ref{subsec_example}.

\subsection{Patterns Without Partial Amplitude Death}
Introducing polar coordinates $z_j=r_je^{i\phi_j}$ in the system of coupled oscillators given by Eqs. (\ref{eq:model}) and (\ref{eq:f}) yields
\begin{equation}
\begin{aligned}
 \dot{r}_i&=(\lambda-r_i^2)r_i+\sigma\sum_jA_{ij}r_j\cos(\beta+\phi_j-\phi_i)~,\\
 r_i\dot{\phi}_i&=\omega r_i+\sigma\sum_jA_{ij}r_j\sin(\beta+\phi_j-\phi_i)~.
\end{aligned}
 \label{eq:polareq}
\end{equation}
We introduce small perturbations, $\delta r_i$ and $\delta\phi_i$, of the limit cycle solution $z_i=v_iz_\eta^{LC}$ in radius and phase
\begin{equation}
\begin{aligned}
 r_i&=r_i^{LC}+\delta r_i=r_\eta+\delta r_i~,\\
 \phi_i&=\phi_i^{LC}+\delta\phi_i=\phi_\eta+\frac{\pi}{2}v_i+\delta\phi_i~.
\end{aligned}
\label{eq:pertub}
\end{equation}
Inserting Eq. (\ref{eq:pertub}) into Eq. (\ref{eq:polareq}) and expanding all right hand sides around $\delta r_i=0,~\delta\phi_i=0$ up to first order yields
\begin{equation}
 \begin{aligned}
 \dot{\delta r_i}=& (\lambda-3r_\eta^2)\,\delta r_i+\sigma\sum_{j=1}^N A_{ij}\left[\cos\left(\beta+\frac{\pi}{2}(v_j-v_i)\right)\delta r_j\right.\\
&\left.-r_\eta\sin\left(\beta+\frac{\pi}{2}(v_j-v_i)\right)\left(\delta\phi_j-\delta\phi_i\right)\right]~,\\[0.1cm]
\dot{\delta\phi_i}r_\eta=&-\eta\sigma\sin\beta\delta r_i+\sigma\sum_{j=1}^NA_{ij}\left[\sin\left(\beta+\frac{\pi}{2}(v_j-v_i)\right)\delta r_j\right.\\
&\left.+r_\eta\cos\left(\beta+\frac{\pi}{2}(v_j-v_i)\right)\left(\delta\phi_j-\delta\phi_i\right)\right]~.
 \end{aligned}
 \label{eq:bla}
\end{equation}
Using $v_i,v_j\in\{-1,+1\}$ and therefore $\frac{\pi}{2}(v_j-v_i)\in\{-\pi,0,\pi\}$ one can write
\begin{align*}
\cos\left(\beta+\frac{\pi}{2}(v_j-v_i)\right)&=v_iv_j\cos\beta~,\\[0.2cm]
\sin\left(\beta+\frac{\pi}{2}(v_j-v_i)\right)&=v_iv_j\sin\beta~.
\end{align*}
Making use of $\sum_jA_{ij}v_j=\eta v_i$, as well as $v_i^2=1$ and additionally introducing $\xi_i=(\delta r_i,\delta\phi_i)^T$, the Eqs. (\ref{eq:bla}) can be rewritten as
\begin{equation}\begin{aligned}
\dot{\xi_i}&=\underbrace{\left(\begin{array}{cc}\lambda-3r_\eta^2 & r_\eta\sigma\eta\sin\beta\\-\frac{1}{r_\eta}\sigma\eta\sin\beta & -\sigma\eta\cos\beta\end{array}\right)}_{J}\xi_i\\
&+\sigma\sum_{j=1}^Nv_iA_{ij}v_j\underbrace{\left(\begin{array}{cc}\cos\beta & -r_\eta\sin\beta\\ \frac{1}{r_\eta}\sin\beta & \cos\beta\end{array}\right)}_{R}\xi_j~.
\end{aligned}
\label{eq:linstab}\end{equation}
After introduction of the diagonal matrix $V=diag(\mathbf{v}$) (with components of the eigenvector $\mathbf{v}$ in its main diagonal) and $\xi=(\xi_1,\xi_2,\dots,\xi_N   )^T$, the system of equations takes the form
\begin{equation}
 \dot{\xi}=\underbrace{\left[I_N\otimes J+\sigma(VAV)\otimes R\right]}_{C}\xi~.
\label{eq:generalstab}
\end{equation}

The dynamical eigenvalues of the $2N\times 2N$ matrix $C$ (defined in Eq. (\ref{eq:generalstab})) determine the stability of the eigensolution. Fortunately, one can show that the matrix $VAV$ in the second term of $C$ is diagonalizable and has the same eigenvalues as $A$, see appendix \ref{app:B}.\\

We can therefore transform Eq. (\ref{eq:linstab}) in a way that diagonalizes $VAV$. We denote the transformed perturbations as $\chi_i$ with corresponding dynamics
\begin{equation}
\dot{\chi_i}=(J+\sigma\mu_iR)\chi_i.
\end{equation}
The $i$th topological eigenvalue of $A$ is now denoted as $\mu_i$ (instead of $\eta^i$) in order not to confuse it with the particular eigenvalue of $A$ that was used to obtain the eigensolution and which we will keep referring to as $\eta$. So there is (at least) one index $i\in\{1,\dots,N\}$ for which $\mu_i=\eta$ is true. Without loss of generality one may assume the topological eigenvalues to be sorted such that
\begin{equation}
 \mu_1\leq\mu_2\leq\dots\leq\mu_N~.
 \label{eq:muorder}
\end{equation}
Now in general the real parts of the eigenvalues of the $2\times 2$ matrices
\begin{equation}
M_i=J+\sigma\mu_iR,\quad i\in\{1,\cdots,N\}
\label{eq:mi}
\end{equation}
need to be negative to secure stability of the eigensolution. 
The characteristic equation for the dynamical eigenvalues $\alpha_i$ of this 2$\times$2 matrix is
\begin{equation}
\alpha_i^2-\text{Tr}[M_i]\alpha_i+\text{Det}[M_i]=0~.
\end{equation}
According to the Hurwitz criterion this equation has solutions with negative real part if and only if
\begin{align}
\text{Tr}[M_i]&=-2\big[\lambda+\sigma \cos (\beta ) (2 \eta -\mu_i )\big]<0\,,\label{eq:tr}\\[0.3cm]
\begin{array}{l}\text{Det}[M_i]\\[0.2cm]~\end{array}&\begin{array}{l}= (\eta -\mu_i )\,\sigma\,\big[2
   \lambda\cos (\beta) +\sigma \{\eta  \cos (2 \beta ).\\[0.2cm]+2 \eta -\mu_i\}\big]>0~.\end{array}\label{eq:det}
\end{align}
We demand that the above inequalities Eq. (\ref{eq:tr}) and Eq. (\ref{eq:det}) are fulfilled for all
\begin{equation}
\text{a) }\mu_i\neq\eta,\quad
\text{b) }\sigma\neq0,\quad
\text{c) }\lambda>-\eta\sigma\cos\beta~.
\label{eq:divconds}
\end{equation}
These conditions restrict the stability analysis to a) perturbations perpendicular to the synchronization manifold, b) the case of coupled oscillators and c) the region of parameter space in which the eigensolution exists. If $\eta$ is degenerate, $\mu_i=\eta$ has to be considered since it no longer solely belongs to a perturbation inside the synchronization manifold. In this case there will always be a dynamical eigenvalue which is zero and the linear stability analysis remains inconclusive.

There are thus $2(N-1)$ inequalities given by Eqs. \eqref{eq:tr} and \eqref{eq:det} determining the stability of the eigensolution belonging to the topological eigenvalue $\eta$. They define regimes of stability in the parameter space of $\lambda, \sigma$ and $\beta$.
These inequalities will now be used to analyze the boundaries of stability in the ($\sigma$,$\lambda$) plane (thus for constant $\beta$) using the definitions
\begin{align}
g(\mu_i,\eta,\beta)&=(\mu_i-2\eta)\cos\beta,
\label{eq:g}\\
 h(\mu_i,\eta,\beta)&=\frac{1}{2\cos\beta}\big[\mu_i-\eta(\cos2\beta+2)\big]~.
\label{eq:h}
\end{align}
The condition from Eq. (\ref{eq:tr}) then reads
\begin{equation}
 \lambda>g(\mu_i,\eta,\beta)\sigma\label{eq:gcond}
\end{equation}
whereas inequality Eq. (\ref{eq:det}) yields the four cases given in Table \ref{tab:hconds2}.\\[0.25cm]
\setlength\tabcolsep{0pt}
\setlength\extrarowheight{4pt}
\begin{table}
\begin{tabular}{|c!{\vrule width 0.05cm}c|c|}\hline\rowcolor{dgrey}
 &$\sigma\cos\beta < 0 $& $\sigma\cos\beta>0$\\\noalign{\hrule height 0.05cm}

\cellcolor{dgrey}$ ~~\mu_i>\eta~~$&$~~\lambda>h(\mu_i,\eta,\beta)\sigma~~$&$~~\lambda<h(\mu_i,\eta,\beta)\sigma~~$\\
\hline\cellcolor{dgrey}
$ \mu_i<\eta$&$\lambda<h(\mu_i,\eta,\beta)\sigma$&$\lambda>h(\mu_i,\eta,\beta)\sigma$\\\hline
\end{tabular}
\caption{Inequality Eq. (\ref{eq:det}) expressed in terms of the function $h$ defined in Eq. (\ref{eq:h}) (white cells). Four different cases of parameter values $\sigma$ and $\beta$ have to be distinguished as indicated by the top row and left column (gray cells).}
\label{tab:hconds2}
\end{table}

For constant $\beta$ the two columns in Table \ref{tab:hconds2} distinguish between $\sigma<0$ and $\sigma>0$. Hence in the ($\sigma$,$\lambda$) plane the two cases that are in the same column define a set of lower ($\lambda>h(\mu_i)$) and a set of upper ($\lambda<h(\mu_i)$) boundaries for $\lambda$ that are given by the different $\mu_i> \eta$ (top row) or $\mu_i<\eta$ (bottom row). 

We thus obtain a series of lower and upper bounds of $\lambda$ for positive $\sigma$ and a possibly different series of upper and lower bounds of $\lambda$ for negative $\sigma$. These need to be combined with yet another set of lower bounds for $\lambda$ defined by the inequalities from Eq. (\ref{eq:gcond}) for all the $\mu_i\neq\eta$ and the criterion for the existence of the solution given in Eq. (\ref{eq:divconds}).

These boundaries are a collection of lines in the ($\sigma,\lambda$) plane that are each going through the origin with a slope determined by either $g(\mu_i,\eta,\beta)$, $h(\mu_i,\eta,\beta)$, or $-\eta\cos\beta$. One can find two of these lines that yield the maximum constraint for $\lambda$ in the regime of $\sigma>0$ or $\sigma<0$ such that in each of these regimes just one upper and one lower boundary for $\lambda$ can be given, each only depending on one selected topological eigenvalue $\mu_i$ and fixed $\beta$ and $\eta$. The inequalities limiting the region of stability can be found in the white cells of Table \ref{tab:combitable} for different cases of parameters $\beta$ and $\sigma$ which are given in the gray cells. Dark gray cells apply to all white cells in their column or row while parameter conditions in light gray cells only apply to the white cell right below them. Crosses in the white cells indicate that for this parameter combination the eigensolution is always unstable.

As can be seen by the different columns of Table \ref{tab:combitable}, the stability region of the eigensolution belonging to the topological eigenvalue $\eta$ differs if $\eta$ is the smallest ($\eta=\mu_1$), the largest ($\eta=\mu_N$), or some intermediate eigenvalue ($\eta=\mu_j$, $j\notin\{1,N\})$. Note that there are critical values of the coupling phase $\beta$ for each of these cases defined by the inequalities in the light gray cells.

\setlength\extrarowheight{6pt}
\setlength\tabcolsep{0pt}
\begin{table*}[ht]
\caption{Conditions for the regimes of stability of an eigensolution belonging to $\eta$ in the parameter space of $\lambda$, $\sigma$ and $\beta$. This table only applies to eigensolutions that do not include partial amplitude death. The inequalities limiting the stable region are given in the white cells. Which of these white cells needs to be considered for a fixed set of $\beta$, $\sigma$ and $\eta$ is determined by the conditions in the gray cells. Dark gray cells apply to all white cells in their row and column while light gray cells only apply to the white cell directly below them.}\bigskip
\begin{tabular}{|c!{\vrule width 0.05cm}c!{\vrule width 0.05cm}c!{\vrule width 0.05cm}c|}\hline
\rowcolor{dgrey}
&$\eta=\mu_1$ & $\eta=\mu_N$ & $\eta=\mu_j$\\
\noalign{\hrule height 0.05cm}
\cellcolor{dgrey} &$\cellcolor{lgrey}~~\cos(2\beta)<0~~$&\cellcolor{lgrey}$\cos(2\beta)<\frac{\mu_1-\mu_{N-1}}{\mu_N-\mu_1}$&\cellcolor{lgrey}$\cos(2\beta)<\frac{\mu_1-\mu_{j+1}}{\mu_j-\mu_1}$\\

\cline{2-4}\cellcolor{dgrey}&$\lambda>g(\mu_2)\sigma$&$~~g(\mu_1)\sigma<\lambda<h(\mu_{N-1})\sigma~~$&$~~h(\mu_{j+1})\sigma<\lambda<h(\mu_{j-1})\sigma~~$
\\

\cline{2-4}\cellcolor{dgrey}$\sigma\cos\beta<0$&\cellcolor{lgrey}$\cos(2\beta)>0$&\cellcolor{lgrey}$\cos(2\beta)>\frac{\mu_1-\mu_{N-1}}{\mu_N-\mu_1}$&\cellcolor{lgrey}$~~\frac{\mu_1-\mu_{j+1}}{\mu_j-\mu_1}<\cos(2\beta)<\frac{\mu_1-\mu_{j-1}}{\mu_j-\mu_1}~~$
\\

\cline{2-4}\cellcolor{dgrey}& &\tikzmark{a1}\hspace{3.3cm}\tikzmark{a2}&$~~g(\mu_1)\sigma<\lambda<h(\mu_{j-1})\sigma~~$\\\cline{4-4}

\cellcolor{dgrey}&$\lambda>h(\mu_2)\sigma$&&\cellcolor{lgrey}$\cos(2\beta)>\frac{\mu_1-\mu_{j-1}}{\mu_j-\mu_1}$\\\cline{4-4}

\cellcolor{dgrey}&&\tikzmark{a3}\hspace{3.3cm}\tikzmark{a4}&\notableentry\\\noalign{\hrule height 0.05cm}

\cellcolor{dgrey} & $\cellcolor{lgrey}\cos(2\beta)<\frac{\mu_N-\mu_2}{\mu_1-\mu_N}$ &\cellcolor{lgrey}$\cos(2\beta)<0$&\cellcolor{lgrey}$\cos(2\beta)<\frac{\mu_N-\mu_{j-1}}{\mu_j-\mu_N}$\\

\cline{2-4}
\cellcolor{dgrey}& $~~g(\mu_N)\sigma<\lambda<h(\mu_2)\sigma~~$ &$\lambda>g(\mu_{N-1})\sigma$&$~~h(\mu_{j-1})\sigma<\lambda<h(\mu_{j+1})\sigma~~$\\
\cline{2-4}
\cellcolor{dgrey}$~~\sigma\cos\beta>0~~$& \cellcolor{lgrey}$\cos(2\beta)>\frac{\mu_N-\mu_2}{\mu_1-\mu_N}$ &\cellcolor{lgrey}$\cos(2\beta)>0$&\cellcolor{lgrey}$~~\frac{\mu_N-\mu_{j-1}}{\mu_j-\mu_N}<\cos(2\beta)<\frac{\mu_N-\mu_{j+1}}{\mu_j-\mu_N}~~$\\

\cline{2-4}
\cellcolor{dgrey}&\tikzmark{b1}\hspace{3.1cm}\tikzmark{b2}&&$~~g(\mu_N)\sigma<\lambda<h(\mu_{j+1})\sigma~~$\\\cline{4-4}

\cellcolor{dgrey}&&$\lambda>h(\mu_{N-1})\sigma$&\cellcolor{lgrey}$\cos(2\beta)>\frac{\mu_N-\mu_{j+1}}{\mu_j-\mu_N}$\\\cline{4-4}

\cellcolor{dgrey}&\tikzmark{b3}\hspace{3.1cm}\tikzmark{b4}&&\notableentry\\\hline
\end{tabular}
\connect[0mm]{a1.north west}{a4.south east}
\connect[0mm]{a2.north east}{a3.south west}
\connect[0mm]{b1.north west}{b4.south east}
\connect[0mm]{b2.north east}{b3.south west}
\label{tab:combitable}
\end{table*}

To illustrate Table \ref{tab:combitable} we consider the motif from Fig. \ref{fig:motif519}. The only pattern without partial amplitude death is the completely in-phase synchronized state (CIS), Fig. \ref{fig:expatterns}a).
It belongs to $\eta=\mu_N=1$, the largest topological eigenvalue. According to the second column of Table \ref{tab:combitable} critical values for $\beta$ are determined by
\begin{equation}
\cos(2\beta)=\frac{\mu_1-\mu_{N-1}}{\mu_N-\mu_1}\quad\text{and}\quad\cos(2\beta)=0~.
\end{equation}
The eigenvalues $\mu_i$ of this motif are
\begin{equation}
\mu_1=-\frac{2}{3}\,,~~\mu_2=-\frac{1}{3}\,,~~\mu_3=\mu_4=0\,,~~\mu_5=1~.
\end{equation}
The critical values of $\beta\in\{0,\pi\}$ are thus
\begin{align}
(\beta\approx 0.99~&\vee~\beta\approx2.15)&\Leftrightarrow~~\cos(2\beta)&=-\frac{2}{5}\\
(\beta=\frac{\pi}{4}~&\vee~\beta=\frac{3\pi}{4})&\Leftrightarrow~~\cos(2\beta)&=0~.
\end{align}
Increasing the coupling phase from $\beta=0$ when crossing $\beta=0.99$, a region of stability between $\lambda=g(\mu_1)\sigma$ and $\lambda=h(\mu_{N-1})\sigma$ is created and later destroyed again at $\beta=2.15$ as given by the upper part of the second column in Table \ref{tab:combitable} where $\eta=\mu_N$ and $\sigma\cos\beta<0$.

When the second set of critical values, $\beta=\pi/4$ and $\beta=3\pi/4$, is crossed, the lower boundary of the region of stability changes between $\lambda=g(\mu_{N-1})\sigma$ and $\lambda=h(\mu_{N-1})\sigma$ as indicated by the lower part of the second column in Table \ref{tab:combitable} where $\eta=\mu_N$ and $\sigma\cos\beta<0$. Note that for fixed $\beta$ the rows of Table \ref{tab:combitable} differentiate between positive and negative $\sigma$. Depending on the value of $\beta$ we thus have either just one stable region limited by the axis at $\sigma=0$ and a lower boundary ($\beta<0.99$ or $\beta>2.15$) or we get an additional region on the other side of the $\sigma=0$ axis that has an upper and a lower boundary ($0.99<\beta<2.15$).

\begin{figure}[h]
\hspace{-0.1cm}\adjustbox{valign=t}{\includegraphics[scale=0.482]{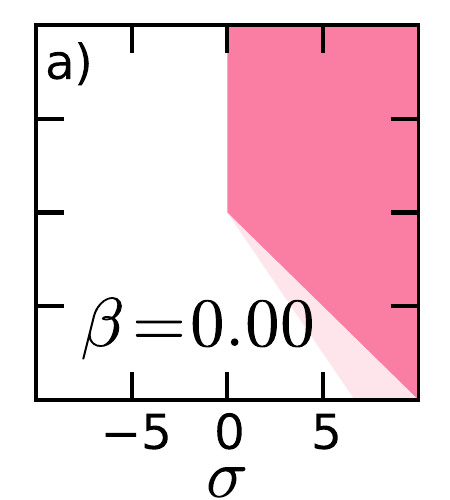}}
\hspace{-0.24cm}\adjustbox{valign=t}{\includegraphics[scale=0.482]{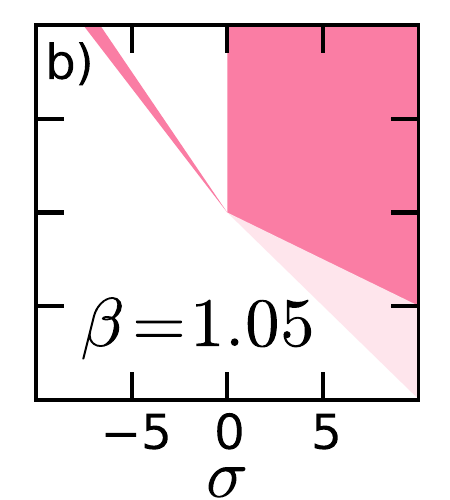}}
\hspace{-0.225cm}\adjustbox{valign=t}{\includegraphics[scale=0.482]{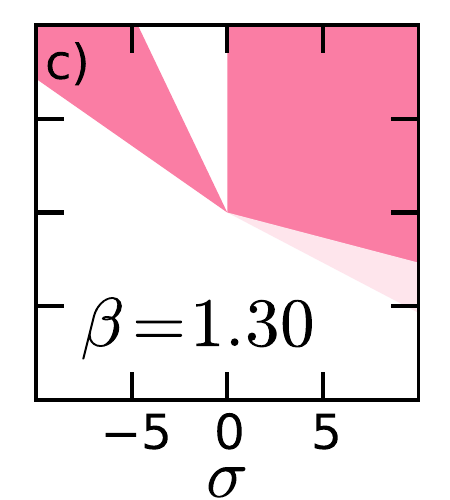}}
\hspace{-0.225cm}\adjustbox{valign=t}{\includegraphics[scale=0.482]{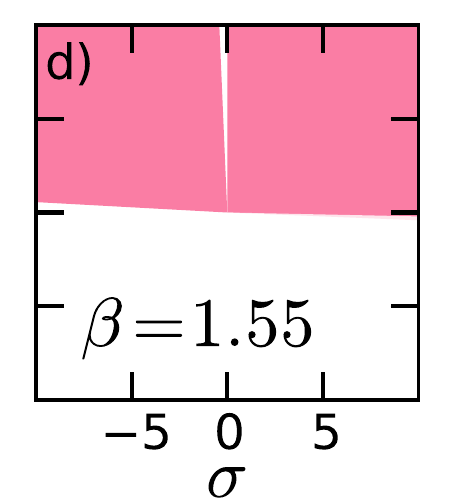}}
\caption{(color online) Stability diagrams for the state of complete in-phase synchronization (CIS, Fig. \ref{fig:expatterns}a)) for the motif shown in Fig. \ref{fig:motif519}. The plots depict the region in the ($\sigma$,$\lambda$) plane in which CIS is stable as colored in red (gray) for a) $\beta=0$, b) $\beta=1.05$, c) $\beta=1.3$, d) $\beta=1.55$. The regions are given by the inequalities in the white cells in Table \ref{tab:combitable}. They are whitened wherever the eigensolution does not exist according to Eq. \eqref{eq:existence}. One clearly sees the appearance of a second stable regime when the critical value of $\beta=0.99$ is crossed.}
\label{fig:cis}
\end{figure}

These regions in which CIS is stable can be seen for different values of $\beta$ in Fig. \ref{fig:cis}. One clearly sees the appearance of an additional regime of stability when the critical value of $\beta=0.99$ is crossed as discussed above.

\subsection{Patterns Including Partial Amplitude Death}\label{subsec_cartesian}
Here we consider patterns including partial amplitude death.
We need to keep cartesian coordinates $z_j=x_j+\mathrm{i}y_j$ to analyze the stability of solutions containing dead oscillators via a linear stability analysis. Using these coordinates equations Eq. (\ref{eq:model}) read
\begin{equation}
\begin{aligned}
\dot{x}_j&=(\lambda-x_j^2-y_j^2)\,x_j-\omega y_j+\sigma\sum_{k=1}^N A_{jk}(\cos\beta x_k-\sin\beta y_k)\,,\\
\dot{y}_j&=(\lambda-x_j^2-y_j^2)\,y_j+\omega x_j+\sigma\sum_{k=1}^N A_{jk}(\sin\beta x_k+\cos\beta y_k)~.
\end{aligned}
\end{equation}
We write the eigensolution as \begin{equation}
z_j(t)=v_jz_\eta(t)=v_jx_\eta(t)+iv_jy_\eta(t)\end{equation}
where
\begin{equation}
\begin{aligned}
x_\eta&=r_\eta\cos\phi_\eta \\
y_\eta&=r_\eta\sin\phi_\eta
\end{aligned}\qquad\text{with}\qquad
\begin{aligned}
 r_\eta&=\sqrt{\tilde{\lambda}}\\
\phi_\eta&=\tilde{\omega}t
\end{aligned}
\end{equation}
and introduce small pertubations $\delta x_j$ and $\delta y_j$\vspace{-0.2cm}
\begin{equation}
\begin{aligned}
x_j&=x_\eta v_j+\delta x_j\\
y_j&=y_\eta v_j+\delta y_j~.
\end{aligned}
\end{equation}
Linearizing around $\delta x_j=0,~\delta y_j=0$ leads to
\begin{equation}
\begin{aligned}
\left(\begin{array}{c}\dot{\delta x_j}\\\dot{\delta y_j}\end{array}\right)&=\bigg[(\lambda-2v_j^2\tilde{\lambda})\,\left(\begin{array}{cc}1 & 0\\0&1\end{array}\right)+\omega\underbrace{\left(\begin{array}{cc}0 & -1\\1&0\end{array}\right)}_{J_1}\bigg]\left(\begin{array}{c}\delta x_j\\\delta y_j\end{array}\right)\\
&-v_j^2\tilde{\lambda}\underbrace{\left(\begin{array}{cc}\cos\left(2\tilde{\omega}t\right) &\sin\left(2\tilde{\omega}t\right)\\\sin\left(2\tilde{\omega}t\right)  &-\cos\left(2\tilde{\omega}t\right)\end{array}\right)}_{J_2(t)}
\left(\begin{array}{c}\delta x_j\\\delta y_j\end{array}\right)\\
&+\sigma\sum_kA_{jk}\underbrace{\left(\begin{array}{cc}\cos\beta &-\sin\beta\\\sin\beta&\cos\beta\end{array}\right)}_{R}\left(\begin{array}{c}\delta x_k\\\delta y_k\end{array}\right)~.\\
\end{aligned}
\label{eq:cartstabfinal}
\end{equation}
 In contrast to the calculations in polar coordinates the Jacobian here depends on time as it contains the matrix $J_2(t)$. To eliminate this dependence we need to change to a co-rotating frame using the rotation matrix
\begin{align}
S(t)=\left(\begin{array}{cc}\cos(\tilde{\omega}t)&\sin(\tilde{\omega}t)\\-\sin(\tilde{\omega}t)&\cos(\tilde{\omega}t)\end{array}\right)~.
\end{align}
In the transformed coordinates
\begin{align}
\left(\begin{array}{c}\delta\bar{x_i}\\
\delta\bar{y_i}
\end{array}\right)=S\left(\begin{array}{c}\delta{x_i}\\\delta{y_i}\end{array}\right)
\end{align}
and after introducing $\xi_j=(\delta\bar{x_j},~\delta\bar{y_j})^T$ and $\xi=(\xi_1,\xi_2,\dots,\xi_n)^T$ Eq. \eqref{eq:cartstabfinal} reads
\begin{equation}\begin{aligned}
\dot{\xi}=\underbrace{\left[I_N\otimes J_A-\tilde{\lambda}V^2\otimes J_B+\sigma A\otimes R\right]}_{C}\xi~,
\label{eq:ccart}
\end{aligned}\vspace{-0.25cm}\end{equation}
with $V=diag(v_i)$ and
\begin{equation}
J_A=\left(\begin{array}{cc}\lambda&\eta\sigma\sin\beta\\-\eta\sigma\sin\beta&\lambda\end{array}\right)~,\quad
J_B=\left(\begin{array}{cc}3&0\\0&1\end{array}\right)~.
\end{equation}
Unfortunately, in general $V^2$ and $A$ do not commute so that we can not reduce the eigenvalue problem as we did in the previous subsection. The eigenvalues of $C$ whose real part, the Lyapunov exponent, determines the stability of the eigensolution therefore need to be calculated numerically for the set of parameters one is interested in. The range of $\lambda\in(-10,10)$, $\sigma\in(-10,10)$, $\beta\in(0,\pi)$, $\omega=2$ was systematically scanned for all motifs. The only significantly large stability region of a PAD+PS state found in this scan is the pattern shown in Fig. \ref{fig:expatterns}b) named PAD+PS1. This stable region is colored in green (light gray) in Fig. \ref{fig:stabdiag} for two different values of $\beta$.

\subsection{The Steady State}\label{subsec_tss}
The stability of the trivial steady state, $z_i=0~\forall~i\in\{1,\dots,N\}$, can be understood completely analytically. If we consider it as a special case of a partially dead state, namely the one with $v_i=0~\forall~i$, we can use some results of the previous section. 

We start from Eq. (\ref{eq:cartstabfinal}) where we insert $v_j=0$ to obtain
\begin{align}
 \left(\begin{array}{c}\dot{\delta x_j}\\\dot{\delta y_j}\end{array}\right)&=\underbrace{\left(\begin{array}{cc}\lambda&-\omega\\\omega&\lambda\end{array}\right)}_{J}\left(\begin{array}{c}\delta x_j\\\delta y_j\end{array}\right)+\sigma\sum_{k=1}^N A_{jk}R\left(\begin{array}{c}\delta x_k\\\delta y_k\end{array}\right)
\end{align}
or with $\xi_j=(\delta x_j,\delta y_j)^T$ and $\mathbf{\xi_j}=(\xi_1,\xi_2,\dots,\xi_N)^T$
\begin{align}
 \dot{\mathbf{\xi}}=\underbrace{\left[I_N\otimes J+\sigma A\otimes R\right]}_{C}\mathbf{\xi}~.
 \label{eq:ctss}
\end{align}
We switch to new coordinates that diagonalize $A$, and the tranformed perturbations $\chi_i$ then obey
\begin{align}
 \dot{\chi_i}=\underbrace{\left(\begin{array}{cc}\lambda+\mu_i\sigma\cos\beta&-(\omega+\mu_i\sigma\sin\beta)\\
                    \omega+\mu_i\sigma\sin\beta&\lambda+\mu_i\sigma\cos\beta\end{array}\right)}_{C_i}\chi_i
\end{align}
where $\mu_i$ are the eigenvalues of the adjacency matrix $A$.
We can apply the criteria from Eq. (\ref{eq:tr}) and Eq. (\ref{eq:det})
\begin{align*}
 \text{Tr}[C_i]&=2(\lambda+\mu_i\sigma\cos\beta)<0\,,\\
 \text{Det}[C_i]&=(\lambda+\mu_i\sigma\cos\beta)^2+(\omega+\mu_i\sigma\sin\beta)^2>0~.
\end{align*}
The first condition can be rewritten as
\begin{equation}
 \lambda<-\mu_i\sigma\cos\beta~.
 \label{eq:hjk}
\end{equation}
If this condition is met, then $(\lambda+\mu_i\sigma\cos\beta)^2\neq0$, and thus the second condition is automatically fulfilled as well. 
Which of the eigenvalues $\mu_1\leq\mu_2\leq\dots\leq\mu_n$ imposes the strongest restriction on $\lambda$ coming from Eq. (\ref{eq:hjk}) is determined by the sign of $\sigma\cos\beta$. It is either the largest, $\mu_N=1$, or the smallest, $\mu_1$.
\begin{table}[htp]
\caption{Analytic stability conditions of the trivial steady state for two different conditions on the coupling parameters $\sigma$ and $\beta$.}\bigskip
\begin{tabular}{|l|c|c|}
\hline
 \rowcolor{dgrey}~~parameter condition~~&$\sigma\cos\beta < 0 $& $\sigma\cos\beta>0$\\
\hline
~~stability condition~~&$~~\lambda<-\mu_1\cos\beta\sigma$~~&$~~\lambda<-\cos\beta\sigma$~~\\\hline
\end{tabular}
\label{tab:tss}
\end{table}

Thus the stable regime of the trivial steady state in the ($\sigma$,$\lambda$) plane is bounded by two lines through the origin, one with slope $-\cos\beta$ and one with $-\mu_1\cos\beta$ as given in Table \ref{tab:tss}.

Now remember that the eigensolution to $\mu_i$ exists if $\lambda\geq-\mu_i\sigma\cos\beta$ (compare Eq. \eqref{eq:existence}). Thus the above conditions are fulfilled and the trivial steady state is stable if and only if none of the eigensolutions exists. This stable region is colored in blue (black) in Fig. \ref{fig:stabdiag}.

\subsection{Example}\label{subsec_example}
\begin{figure}[ht]
a)\adjustbox{valign=t}{\includegraphics[scale=0.3]{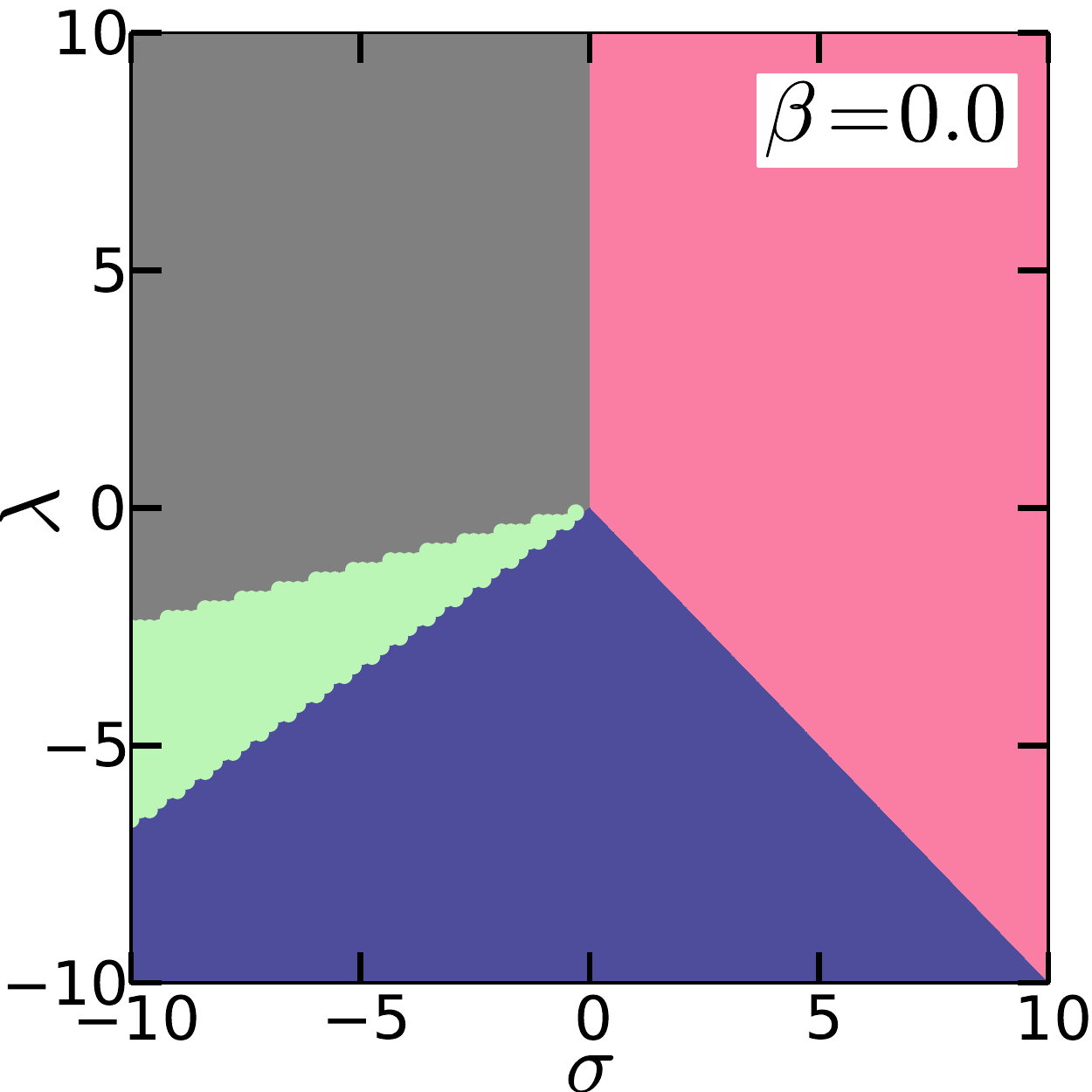}}
b)\adjustbox{valign=t}{\includegraphics[scale=0.3]{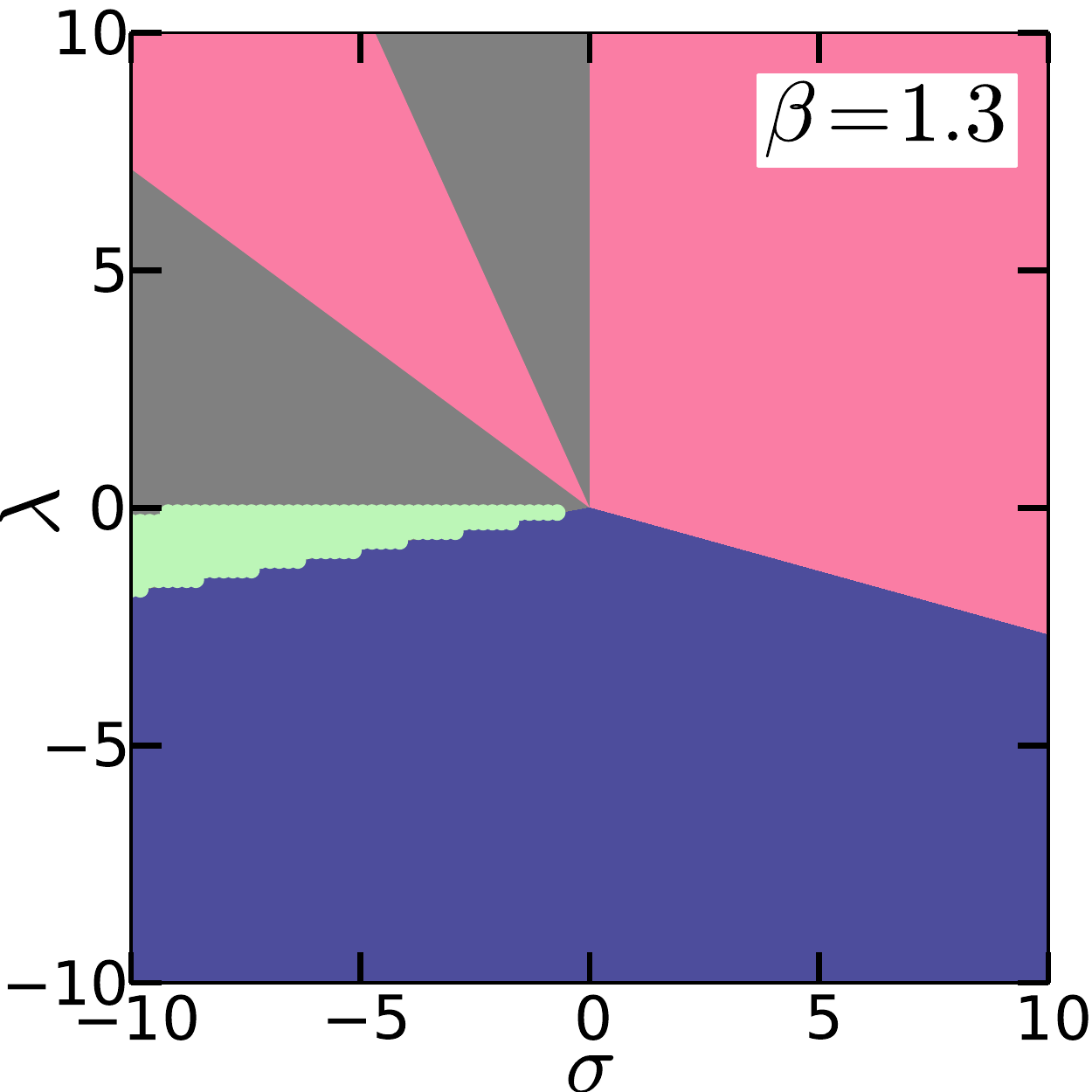}}\\[0.1cm]
\includegraphics[scale=0.25]{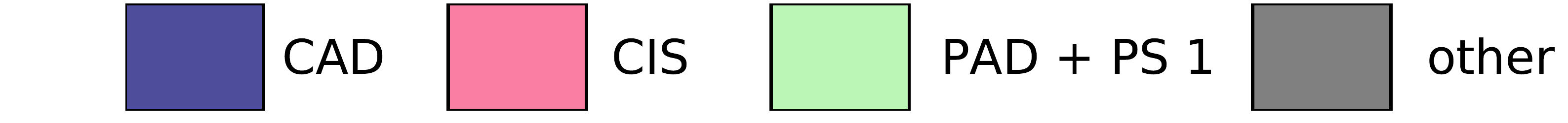}
\caption{(color online) Semi-analytic stability diagram for the homogeneous motif of Stuart-Landau oscillators as shown in Fig. \ref{fig:motif519} for a) $\beta=0$ and b) $\beta=1.3$. The regions are colored according to the pattern from Fig. \ref{fig:expatterns} that is stable there. The dark grey region denoted as `other' indicates that none of the known patterns is stable. The stability of CAD and CIS is known purely analytically from tables \ref{tab:combitable} and \ref{tab:tss}, while the stability of the partially dead solutions PAD+PS 1 to 3 was calculated numerically from Eq. (\ref{eq:ccart}). Of these only PAD+PS 1 is found to be stable.}
\label{fig:stabdiag}
\end{figure}

The results of the above stability analysis are combined for the motif of Fig. \ref{fig:motif519} in the stability diagrams in Fig. \ref{fig:stabdiag} for $\beta=0$ and $\beta=1.3$. The two values of $\beta$ were chosen to represent the case without and with the additional region of stability for CIS (see Fig. \ref{fig:cis}a) and c)). The stability of CAD and CIS are analytical results given in Table \ref{tab:combitable} and \ref{tab:tss} respectively, while the stable region of PAD+PS 1 was obtained by numerically calculating the dynamical eigenvalues from Eq. \eqref{eq:ccart} for the different sets of parameters $\lambda$, $\sigma$ and $\beta$ for $\eta=-2/3$.

\section{Robustness Under Frequency Mismatch} \label{sec_het}
Real world systems are typically subject to heterogeneities in the parameters while in our model the oscillators are homogeneous. To address this problem we numerically test our previous observations for heterogeneous oscillator frequencies.

The system of coupled differential equations obtained from Eq. (\ref{eq:model}) by introducing individual oscillator frequencies $\omega_i$ is
\begin{equation}
\begin{aligned}
\dot{z_i}&=f_i(z_i)+\sigma e^{i\beta}\sum_{j=1}^N A_{ij}z_j\\
f_i(z)&=(\lambda+i\omega_i-|z|^2)z~.
\end{aligned}
\label{eq:problemhet}
\end{equation}
The previously introduced analytical method fails for heterogeneous frequencies $\omega_i$.
For the numerical investigation of coupled Stuart-Landau oscillators the ordinary differential equations given in Eq. (\ref{eq:problemhet}) that describe the system's dynamics are integrated numerically until transients have vanished and the system assumes an asymptotic state. The last few periods of the obtained time series are then used to classify the asymptotic state reached for the chosen parameter values of $\lambda$, $\sigma$, $\beta$. 

Ideally, one would systematically scan the entire parameter space, but here we will focus on the cases of $\beta=0$ and $\beta=1.3$ in the coupling phase to keep the numerical expense reasonable for this study. These cases refer to a  real coupling ($\beta=0$) and an example of complex coupling ($\beta=1.3$) for which the state of complete in-phase synchronization is known to have an additional region of stability as can be seen in Fig. \ref{fig:stabdiag} b) for the homogeneous system. For $\sigma$ and $\lambda$ the parameter space is scanned systematically within the range $(-10, 10)$. This range may even be sufficient since the stability analysis in Sect. \ref{sec_stabana} shows that knowledge of the system close to the origin explains stability for all other values of $\lambda$ and $\sigma$.

Because there might be multistability in our system we have to choose initial conditions carefully. We are looking for the eigensolutions sketched in Fig. \ref{fig:expatterns} and thus use states close to them  as initial conditions. To do so, one snapshot in time of the analytic solution for $z_i(t)$ is taken and a random signal with entries of the order of 5\% of the solution's radius is added. Additionally random initial conditions are tested, where the real and imaginary part of each oscillator are chosen randomly from the interval (-1, 1).

Once the numerical integration is finished, the very last part of the time series is used to classify the behavior. The different categories are:
\begin{itemize}\label{thispage}
\item \textbf{Complete Amplitude Death} (CAD): All oscillators have a radius equal to zero.
\item \textbf{Complete In-phase Synchronization} (CIS): All oscillators have the same constant finite radius and are inphase synchronized.
\item \textbf{Partial Amplitude Death and Partial Synchronization} (PAD + PS): Some of the oscillators have a radius equal to zero and the others all have the same constant finite radius and are anti-phase or in-phase synchronized. The spatial distribution of these groups is according to one of the patterns in Fig. \ref{fig:expatterns}. There is thus one category for each of the patterns that include partial amplitude death. They are labeled as shown in Fig. \ref{fig:expatterns}.
\item \textbf{Partial Amplitude Death} (PAD): Some of the oscillators have a radius equal to zero. The others are not synchronized as in one of the previously defined patterns.
\end{itemize}
The algorithm checks the last 50 out of 8000 time steps of the simulation for these properties as follows. A critical threshold $r_{crit}$ is defined as 1\% of the maximum radius occurring in all the oscillators in this time interval. If this is smaller than 10 times an internal numerical threshold, the machine epsilon \footnote{This is an upper bound to the error that may occur due to rounding in numerical floating point arithmetic and is given by the NumPy package used for the algorithm as $\epsilon=2.2204460492503131\cdot 10^{-16}$}, the state is considered to be CAD. If not, each oscillator is classified by its radius. A radius smaller than $r_{crit}$ is considered zero and a signal with $r_{min}\geq0.98r_{max}$ and $r_{max}>r_{crit}$ is identified as a constant finite radius.  If the mean radii of all oscillators that are not dead do not differ among each other for more than $r_{crit}$, these oscillators are regarded to have the same radius.

As a measure for synchronization among the active ($r\neq0$) oscillators a slightly modified version of the Kuramoto order parameter \cite{KUR84} is used. The Kuramoto order parameter is equal to unity for complete synchronization and zero for complete desynchronization. It is commonly defined as
\begin{equation}
R=\frac{1}{N}\left|\sum_j^{N}e^{i\phi_j}\right|~,
\end{equation}
where N is the number of oscillators whose synchronization behavior one is interested in. We introduce a slight change that takes into account the fixed phase differences of the oscillators in the investigated patterns and can define a similar order parameter for the synchronization pattern belonging to the eigensolution given by $\mathbf{v^k}$
\begin{equation}
R^k=\frac{1}{N_a}\left|\sum_{j=1}^N v_j^ke^{i\phi_j}\right|~,
\label{eq:rk}
\end{equation}
where $N_a$ is the number of active oscillators in the pattern, $N_a=\sum_j|v_j|$, and $v_j^k$ is the $j$-th component of the $k$-th eigenvector of $A$. The parameter $R^k$ becomes unity if the oscillators show the pattern of in-phase and anti-phase synchronization defined by $\mathbf{v^k}$. We consider the Stuart-Landau oscillators synchronized if  $R\geq0.99$ over the last 50 time steps.

\begin{figure}[ht]
a)\adjustbox{valign=t}{\includegraphics[scale=0.3]{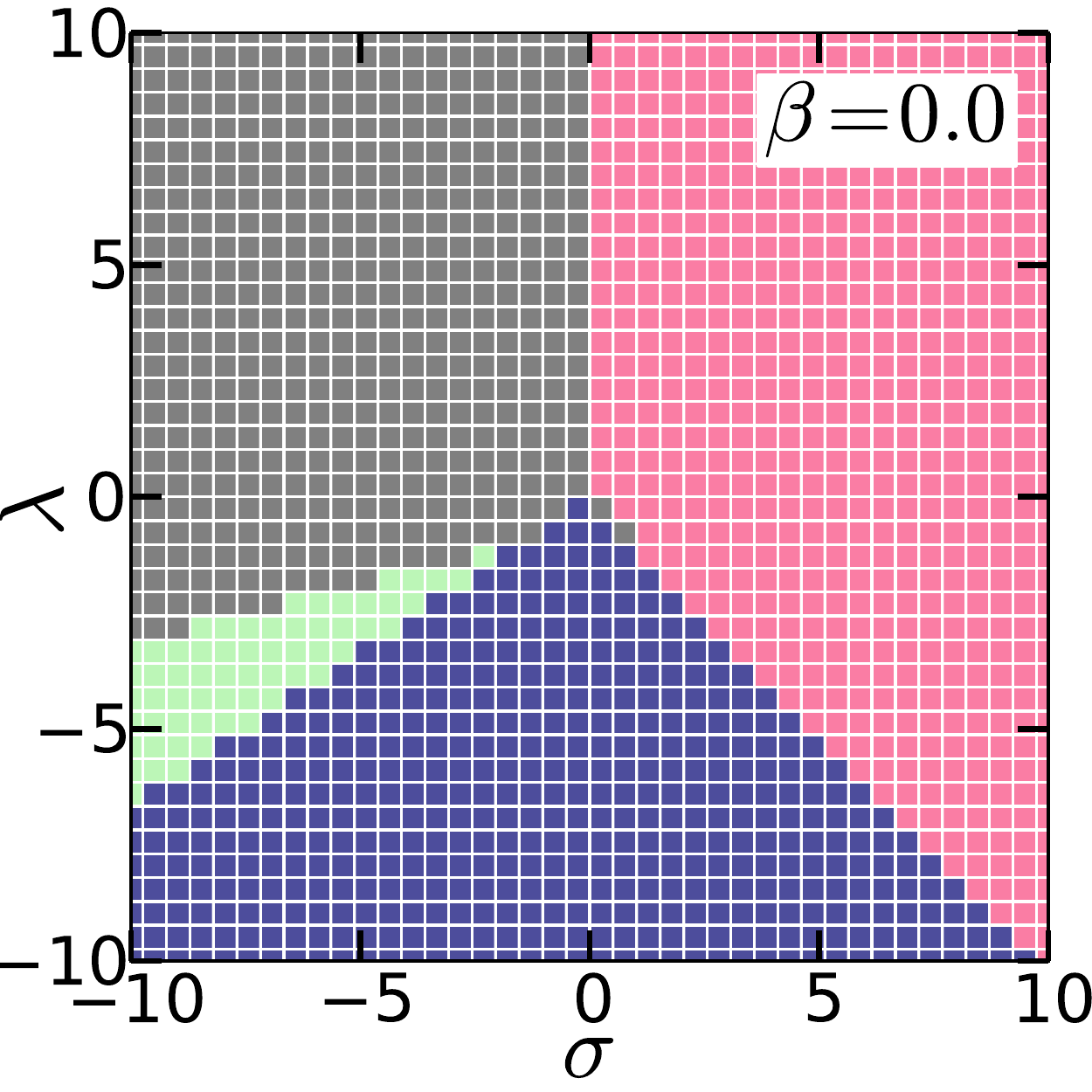}}
b)\adjustbox{valign=t}{\includegraphics[scale=0.3]{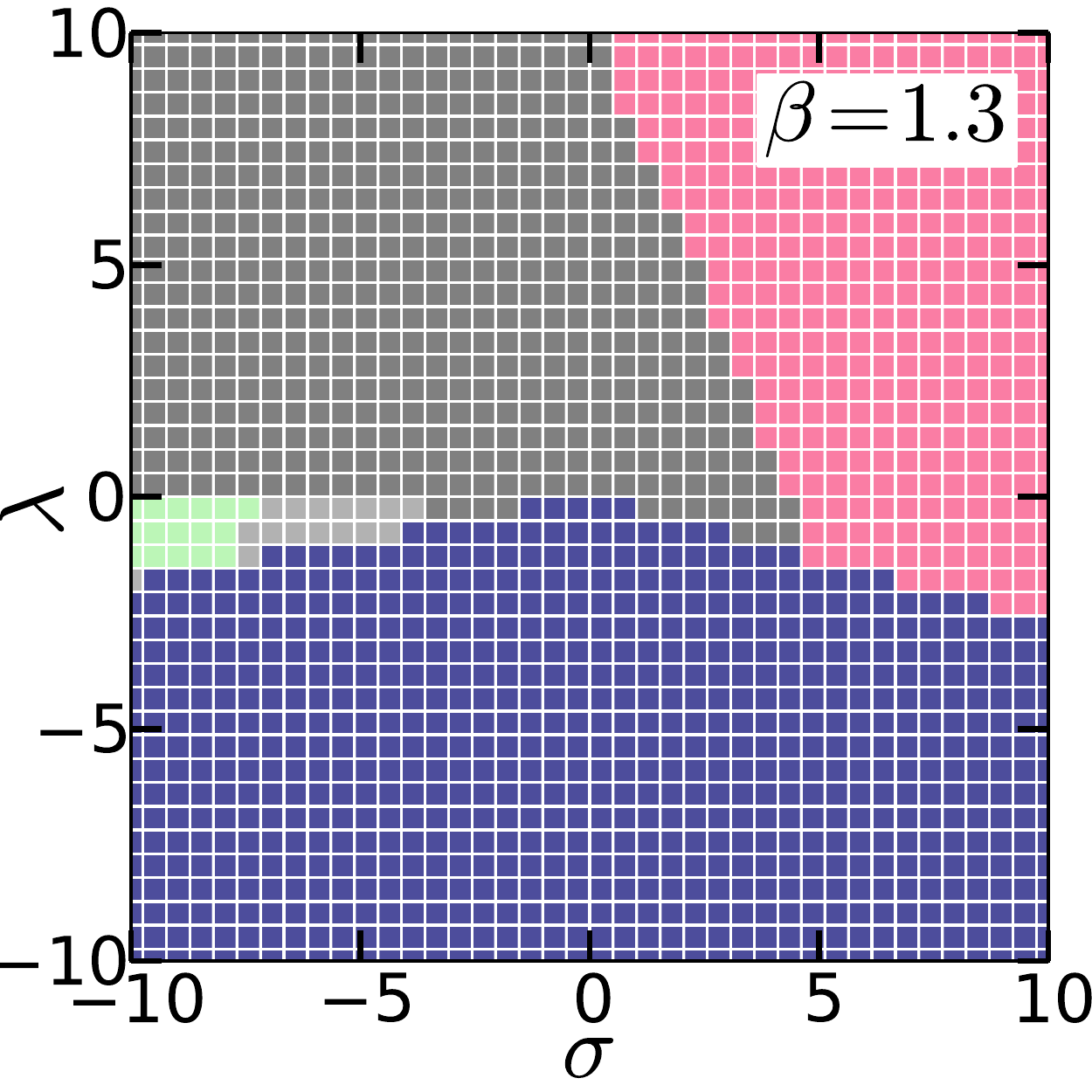}}\\
~~\adjustbox{valign=t}{\includegraphics[scale=0.3]{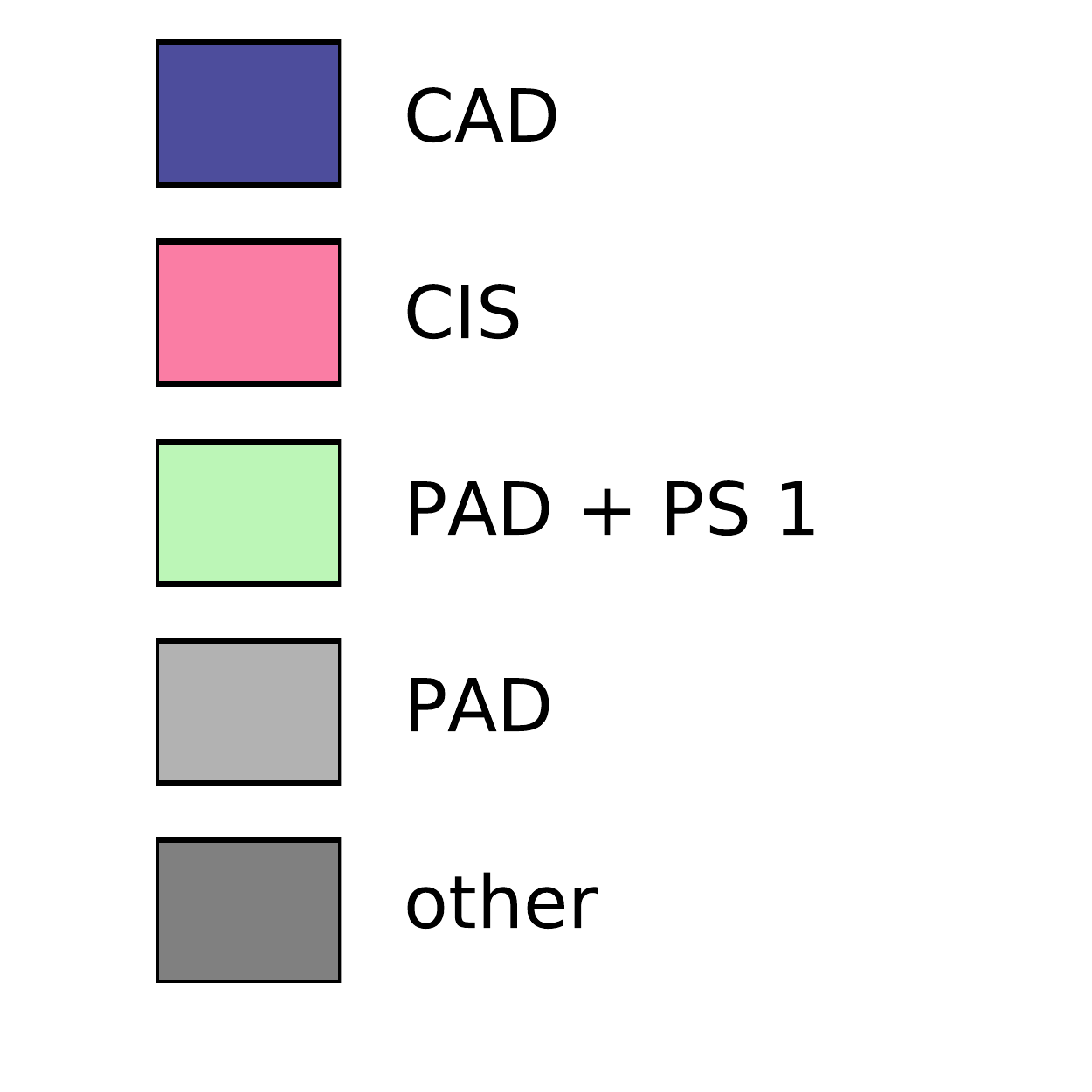}}
c)\adjustbox{valign=t}{\includegraphics[scale=0.3]{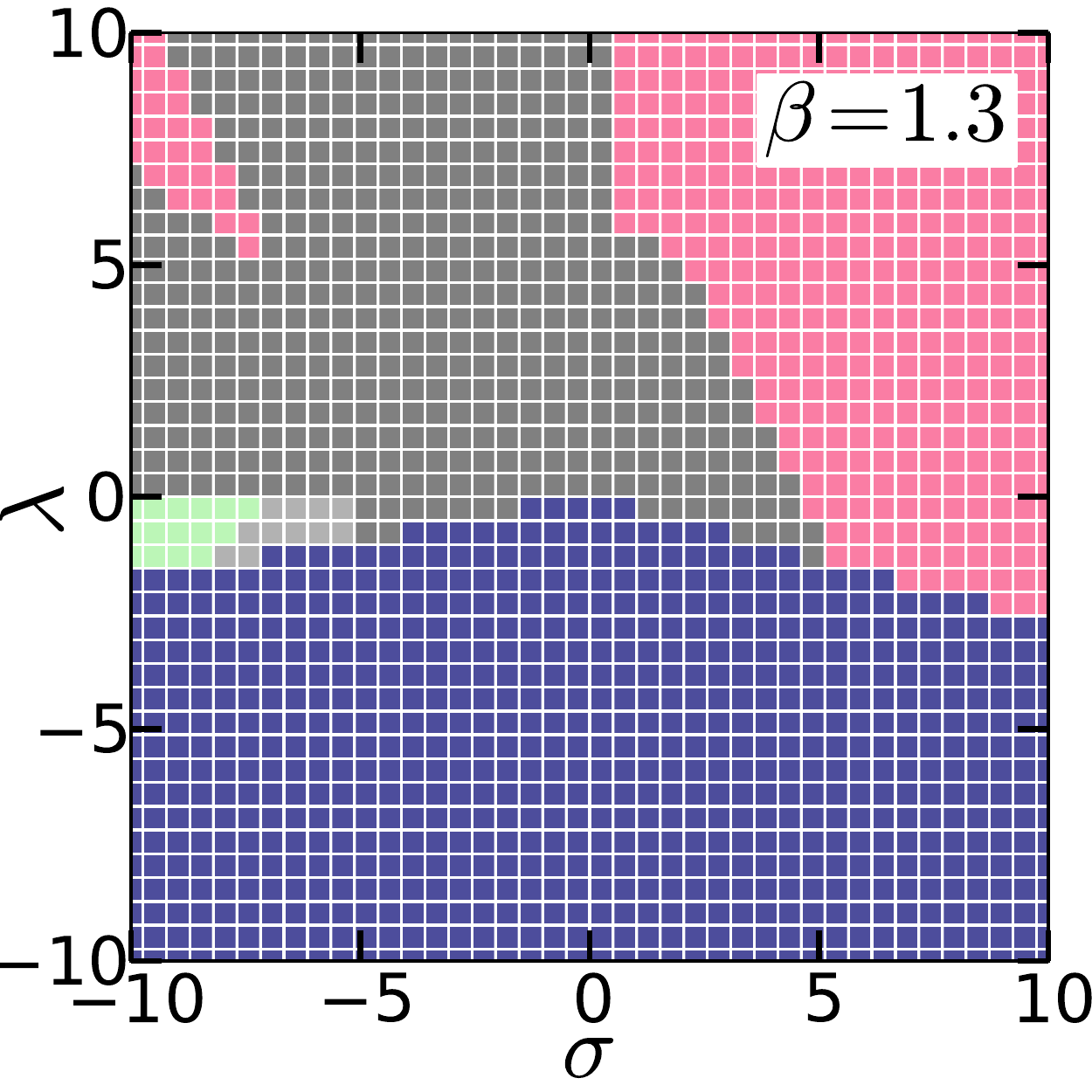}}
\caption{(color online) Numerical stability diagrams for the heterogeneous motif of Stuart-Landau oscillators as shown in Fig. \ref{fig:motif519}. Panels a) and b) have random initial conditions while panel c) was initially close to the CIS state. Each panel has its own set of random frequencies $\omega_i=\omega\cdot\text{rand}(0.95,1.05),~\omega=2$. The color of each point indicates the state that has been reached after numerically integrating until $t=8000$. These panels differ from the analytic stability diagram of the homogeneous system, Fig. \ref{fig:stabdiag}, in the size of the PAD+PS 1 region and the shape of the CIS region. One can also see multistability in the additional CIS region for $\beta=1.3$.}
\label{fig:het}
\end{figure}

The plots obtained for random initial conditions are displayed in Fig. \ref{fig:het} for $\beta=0$ and $\beta=1.3$ together with the plot with initial conditions close to the CIS state for $\beta=1.3$, which shows the additional stable CIS region already seen in Fig. \ref{fig:stabdiag} for homogeneous oscillators. All other initial conditions yield results similar to panels a) and b) in Fig. \ref{fig:het}. Thus, except for the additional CIS region for $\beta=1.3$, the eigensolutions do not seem to have any coexisting stable attractors. 

The numerical stability diagrams for the heterogeneous system exhibit a structure that is very similar to the analytic diagrams for the homogeneous case depicted in Fig. \ref{fig:stabdiag}. All categories are found in both diagrams and in roughly the same region of parameter space. The first main differences lies in the PAD+PS 1 region. It is slightly smaller for the heterogeneous than for the homogeneous system and is surrounded by a region in which PAD without any special type of synchronized behavior is present. This suggests that even though PAD may be robust to frequency mismatch, the partial synchronization may not. On the other hand it is also possible that in the heterogeneous system very long transients are encountered and that thus the PAD+PS 1 state may still be reached after longer integration times in the PAD region.

The other main difference, the change in the CIS region, is only seen for $\beta=1.3$. For $\beta=0$ the border of the CIS and CAD state shows some states classified as `other' but these are most likely due to very long transients at the border of stability where the Lyapunov exponent is close to zero. For $\beta=1.3$ though the shape of the CIS region changes significantly and in a similar way for all initial conditions and randomly drawn sets of frequencies. Thus this effect might be a general result of the heterogeneity and not be due to merely numerical issues as the measure of synchronization or long transients. The new shape suggests that with heterogeneity a critical coupling strength (depending on $\lambda$) has to be reached before the state of complete in-phase synchronization is stable.  This seems plausible since the uncoupled oscillators all have different frequencies and would never synchronize unless forced by their interaction.

\section{Towards Complex Networks}\label{sec_tcn}
Here we present first steps in extending the concept of eigensolutions to larger networks.

For this purpose two approaches are chosen. First, we will extend the network by connecting motifs or adding a dead node, and second, we will identify special eigenvectors that are present independently of system size for the coupling matrix $A$ \footnote{See Supplemental Material at [] for proofs of the results presented in Sect. \ref{sec_tcn}}.

\subsection{Connecting Motifs Via a Dead Node}
\begin{figure}[htp]
\adjustbox{valign=c}{\includegraphics[scale=0.23]{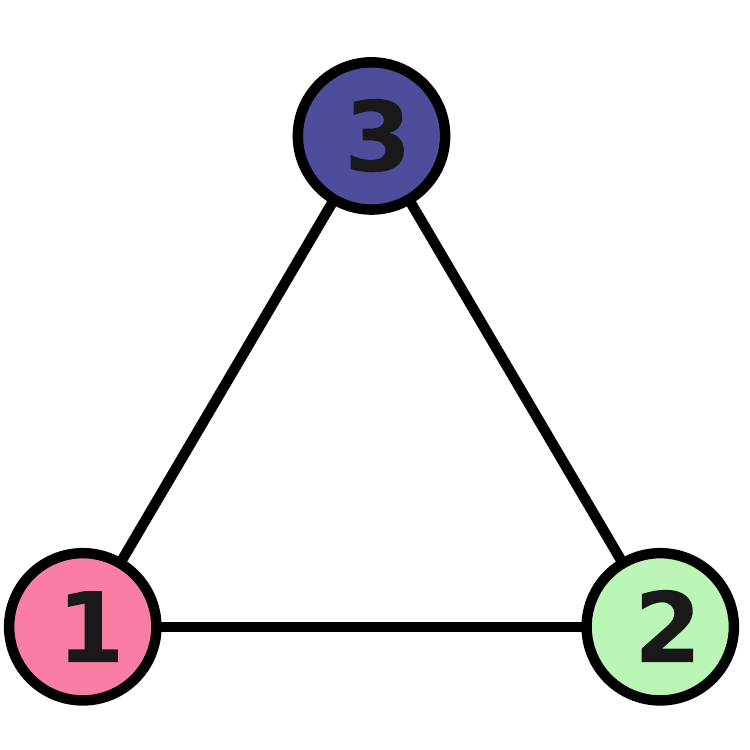}}{\large+}
\adjustbox{valign=c}{\includegraphics[scale=0.23]{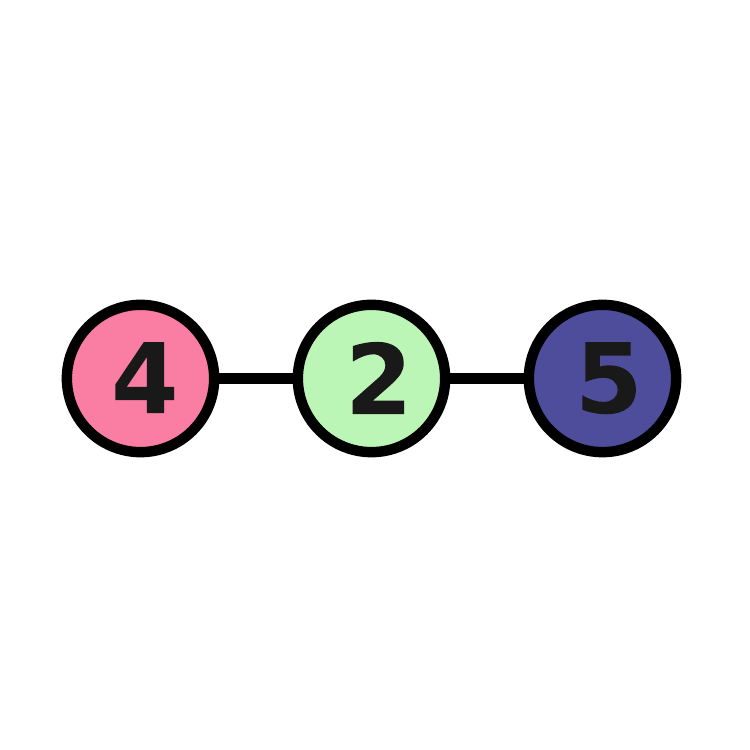}}{\large~~=~~~}
\adjustbox{valign=c}{\includegraphics[scale=0.23]{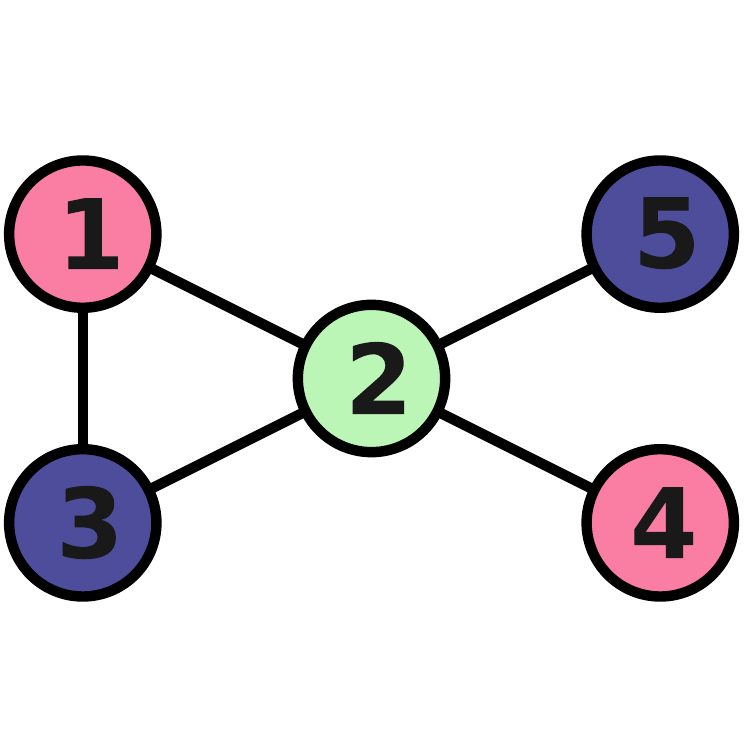}}
\caption{(color online) Two motifs with partially dead eigensolutions can be connected at their dead nodes without changing their patterns. This may create patterns with more than one frequency.}
\label{fig:connection}
\end{figure}

If two motifs each have an eigensolution with a dead node these nodes can be combined to form a larger network as illustrated in Fig. \ref{fig:connection}. The combined network then has a spatio-temporal pattern given by the two motifs' eigensolutions. In formulas this reads as follows.

Let the two single motifs have eigensolutions
\begin{align*}
\mathbf{z^1}=z_\eta\mathbf{v^\eta}\in\mathbb{R}^M~,\\
\mathbf{z^2}=z_\mu\mathbf{v^\mu}\in\mathbb{R}^N
\end{align*}
containing each a dead node. Without loss of generality we can fix them as $v^\eta_M=0$ and $v^\mu_1=0$.
Then the resulting larger motif with $M+N-1$ nodes has the solution $\mathbf{z}\in\mathbf{R}^{M+N-1}$ with entries
\begin{equation}
\begin{aligned}
z_i=\left\{\begin{array}{cl}z_\eta v^\eta_i& \text{if~~} i\in\{1,\dots,M-1\}\\
0 &\text{if~~} i=M\\
z_\mu v^\mu_{i-(M-1)}&\text{if~~} i\in\{M+1,\dots,M+N-1\}\end{array}\right.~.
\end{aligned}
\end{equation}
This is also true if one or both of the eigenvectors are replaced by the zero-vector of the same dimension. Thus one can attach an arbitrarily shaped network to an eigensolution's dead node, and still the eigensolution remains valid on the small motif while the attached network is in the trivial steady state.

If the two eigensolutions belong to different eigenvalues, $\eta\neq\mu$, the combined solution will have more than one frequency and amplitude because of 
\begin{equation}
\begin{aligned}
 z_\eta&=\sqrt{\lambda+\eta\sigma\cos\beta}e^{i(\omega+\eta\sigma\sin\beta)t}\\
 z_\mu&=\sqrt{\lambda+\mu\sigma\cos\beta}e^{i(\omega+\mu\sigma\sin\beta)t}~.
\end{aligned}
\end{equation}

For example, in the case shown in Fig. \ref{fig:connection}, the left (triangular) motif's eigensolution belongs to $\eta=-1/2$ while the right (linear) motif's eigensolution belongs to $\mu=0$. Thus in the combined motif the oscillators 1 and 2 coming originally from the triangular motif have a different frequency and amplitude than oscillators 4 and 5 coming originally from the linear motif
\begin{align*}
 z_1&=z_2=\sqrt{\lambda-\frac{1}{2}\sigma\cos\beta}\,e^{i(\omega-\frac{1}{2}\sigma\sin\beta)t}~,\\
z_3&=0~,\\
 z_4&=z_5\sqrt{\lambda}\,e^{i\omega t}~.
\end{align*}
This combined pattern is no longer based on a single eigenvector of the combined networks adjacency matrix, but rather on a linear combination of eigenvectors.

\subsection{Attaching a Dead Node to a Set of Active Nodes}
\begin{figure}[htp]
a)\hspace{0.1cm}\adjustbox{valign=t}{\includegraphics[scale=0.23]{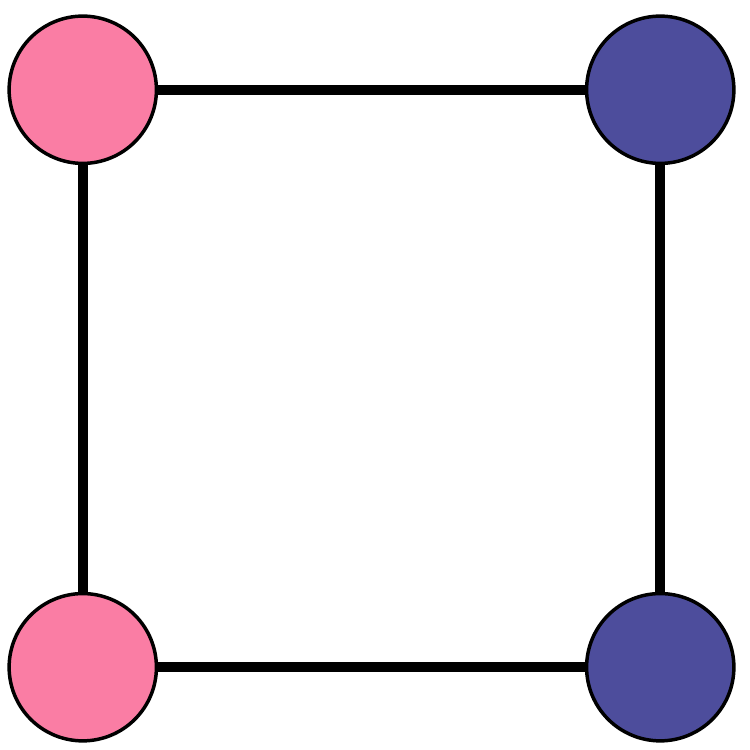}}\hspace{0.2cm}
b)\hspace{0.1cm}\adjustbox{valign=t}{\includegraphics[scale=0.23]{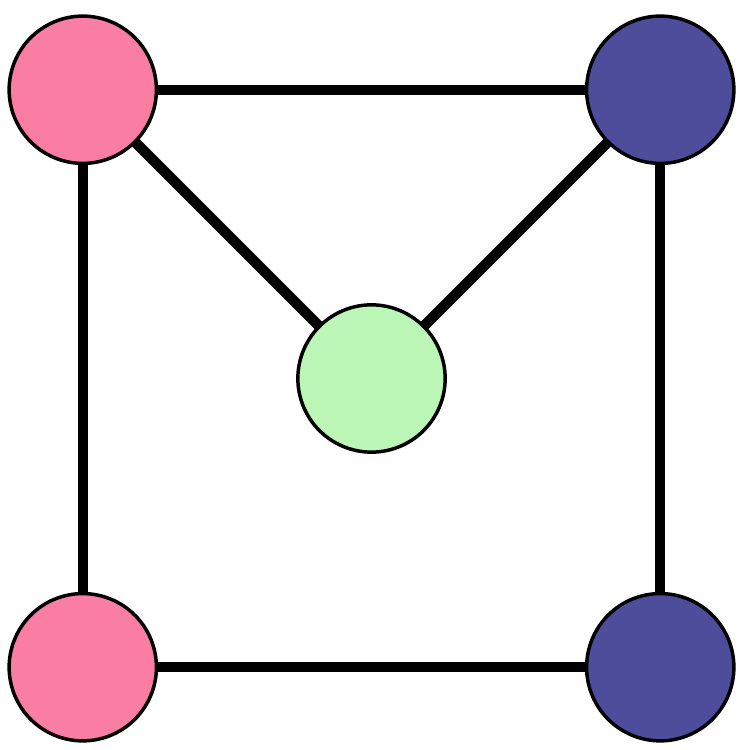}}\hspace{0.2cm}
c)\hspace{0.1cm}\adjustbox{valign=t}{\includegraphics[scale=0.23]{Motif519c2}}
\caption{(color online) A dead node (green/light gray) can be attached to a network without changing an eigensolution belonging to $\eta=0$ if all neighbors of the dead node sum up to zero. In this example the dead node is attached to the 4-node motif with partially synchronized eigensolution shown in a). There are two possible ways to do this which are depicted in b) and c). }
\label{fig:atdead}
\end{figure}

If an eigensolution corresponds to $\eta=0$, one can attach a dead oscillator by connecting it such that its neighbors sum up to zero. If we call the newly created network's adjacency matrix (containing the additional dead node) $B$ this translates to
\begin{equation*}
\sum_j B_{ij}z_j=0.
\end{equation*}
For other values of $\eta$ this does not work. Some examples can be seen in Fig. \ref{fig:atdead} where a dead node is attached in two different manners to a 4-node ring.

Together these two mechanism make it possible to create large networks containing synchronization clusters with different frequencies and partial amplitude death. Given an arbitrary network one can always place as many motif eigensolutions on it as possible under the constraint of connecting them via dead nodes. All the other nodes can then be set to zero and thus one has constructed a solution of the large arbitrary network.

\subsection{Eigensolutions Existing For All Network Sizes}
As pointed out in Sect. \ref{sec_model} the 
 CIS state exists for all network topologies independently of size because of our choice of the coupling matrix $A$. Also, one can show that for bipartite graphs the matrix $A$ always has an eigenvector with interchanging entries $v_i=1$ and $v_j=-1$ such that oscillators with $v_i=1$ are only connected to those with $v_j=-1$. This is another partially synchronized eigensolution that is independent of the network size and it always corresponds to the topological eigenvalue $\eta=-1$. 

Our analytic stability analysis from Sect. \ref{sec_stabana} holds for all system sizes. According to Table \ref{tab:combitable} the stability of the CIS state ($\eta=\mu_N=1$) only depends on the second largest, $\mu_{N-1}$ and the smallest topological eigenvalue, $\mu_1$. The stability of the PS state on a bipartite network belonging to $\eta=\mu_1=-1$ is governed by the second smallest, $\mu_2$ and the largest, $\mu_N$, topological eigenvalue. Since by construction of $A$ we always get $\mu_N=1$ the stability really only depends on $\mu_2$ in this case.

Thus our work analytically describes two general types of behavior for an arbitrarily large number of homogeneous Stuart-Landau oscillators coupled via a normalized adjacency matrix. One could possibly think of an application where the CIS state can be (de)stabilized by adding or removing certain nodes or edges as observed, e.g., for power grids \cite{WIT12}. In order to do this one would not need to compute the entire spectrum of topological eigenvalues but only the second largest and the smallest. 
\begin{figure}[hpt]
a)\hspace{0.1cm}\adjustbox{valign=t}{\includegraphics[scale=0.2]{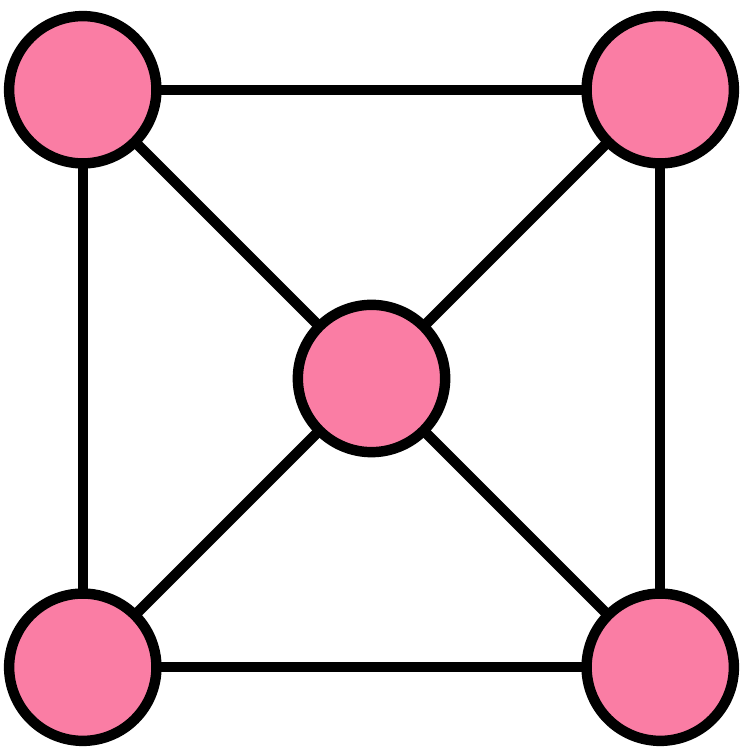}}
\hspace{0.2cm}b)\hspace{0.1cm}\adjustbox{valign=t}{\includegraphics[scale=0.2]{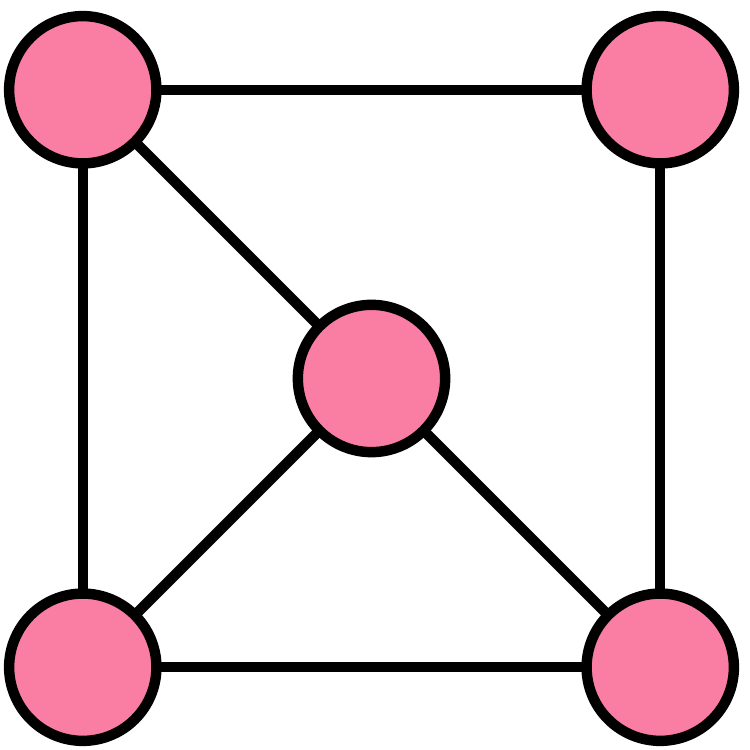}}
\hspace{0.2cm}c)\hspace{0.1cm}\adjustbox{valign=t}{\includegraphics[angle=180,origin=c,scale=0.2]{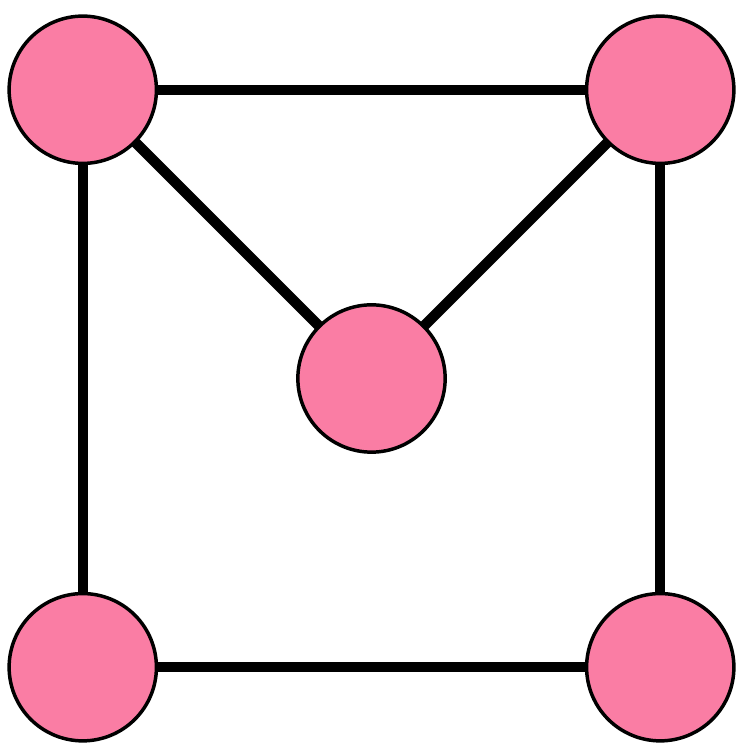}}
\hspace{0.2cm}d)\hspace{0.1cm}\adjustbox{valign=t}{\includegraphics[scale=0.2]{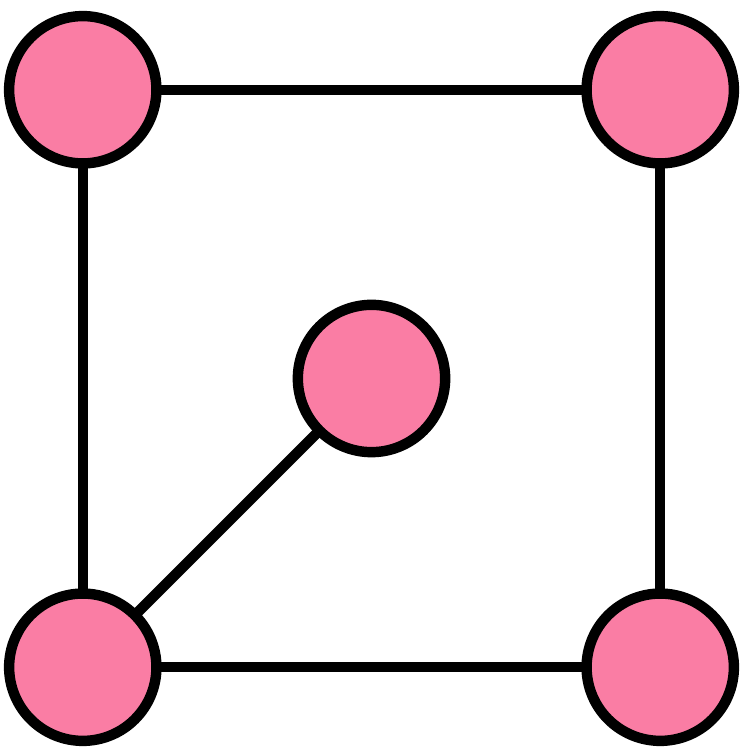}}
\caption{(color online) The topology of the motif from Fig. \ref{fig:motif519} is changed step by step by detaching the middle node. This process changes the topological eigenvalues (Table \ref{tab:cisex}) and thus the stability of the CIS state as can be seen in Fig. \ref{fig:cisexample}.}
\label{fig:cisex}
\end{figure}
An example of how one may change the topology to influence the stability of CIS is shown in Fig. \ref{fig:cisex}. The previously discussed five-node motif is altered step by step by slowly detaching the middle node. The change in the smallest and second largest eigenvalue introduced this way can be seen in Table \ref{tab:cisex}. The different regions of stability defined by these topological eigenvalues according to the upper part of the second column in Table \ref{tab:combitable} are presented in Fig. \ref{fig:cisexample}.

\setlength\tabcolsep{0pt}
\setlength\extrarowheight{2pt}
\begin{table}[ht]
\caption{Smallest, $\mu_1$, and second largest, $\mu_{N-1}$, topological eigenvalue of the motifs shown in Fig. \ref{fig:cisex}. These eigenvalues determine the stability of the CIS state that can be seen in Fig. \ref{fig:cisexample}.}\bigskip
\begin{tabular}{|c!{\vrule width 0.05cm}c|c|c|c|}\hline\rowcolor{dgrey}
&~a)~ &~b)~ &~c)~ &~d)~~~\\\noalign{\hrule height 0.05cm}
\cellcolor{dgrey}$~\mu_{N-1}~$ & ~0~ & $~\frac{1}{3}(-1+\sqrt{2})$~ &~$\frac{1}{3}$~  &$~\frac{1}{\sqrt{6}}~$\\\hline
\cellcolor{dgrey}$~\mu_1~$ & $~-\frac{2}{3}$~ &$~-\frac{1}{3}(1+\sqrt{2})$~ &~$-\frac{5}{6}~$  &-~1~\\\hline
\end{tabular}
\label{tab:cisex}
\end{table}
\begin{figure}[hpt]
\includegraphics[scale=0.6]{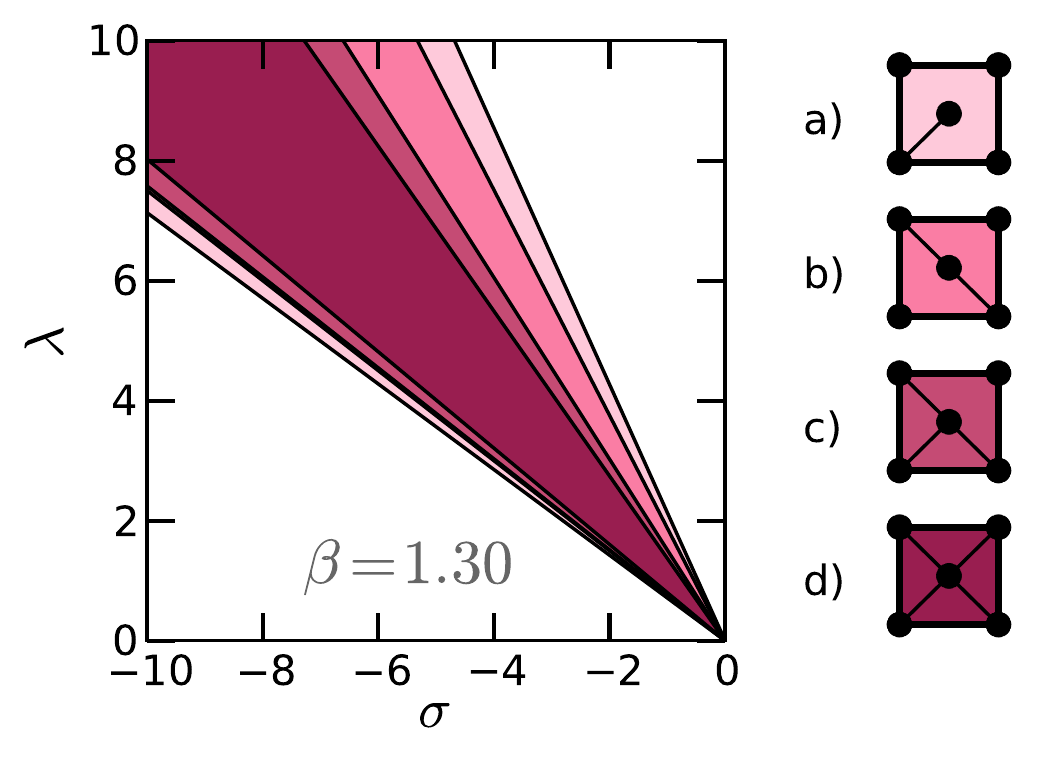}
\caption{(color online) Analytic stability diagram for the motifs shown in Fig. \ref{fig:cisex} a) to d). The CIS state is stable in the region shaded according to the legend on the right. There is just one connected region for each color with dark colors overlaying the lighter ones. Changing the network topology clearly changes the stability of the CIS state. The change is governed solely by the change of the smallest and second largest topological eigenvalue given in Table \ref{tab:cisex}.}
\label{fig:cisexample}
\end{figure}
\section{Conclusion} \label{sec_conc}
The main achievement of this paper is the development of eigensolutions as a concept to analytically predict the coexistence of partial in-phase and anti-phase synchronization with partial amplitude death. The connection between topology and dynamics becomes clearly visible and analytically accessible in this concept. One could use eigensolutions to derive dynamical patterns of 
partial amplitude death and partial synchronization solely from the network topology. This, however, requires a thorough understanding of the special eigenvectors needed for its construction, especially of the conditions under which they are present in a network. Therefore investigating the occurrence of these eigenvectors is the next important step that should be taken in future work.

The stability of eigensolutions without partial amplitude death was determined to be dependent directly on a small subset of the network's topological eigenvalues, Sect. \ref{sec_stabana}. After the introduction of eigensolutions this is the second strong conclusion about the direct interplay of dynamics and topology made in this work. For the general case of eigensolutions containing partial amplitude death and partial synchronization the stability  for a set of parameters can be calculated numerically from the analytically obtained time independent Jacobian. This Jacobian should be further analyzed and may yield analytic stability conditions in the form of inequalities as well.

The prediction of dynamical patterns via eigensolutions has proved not to be restricted to networks of homogeneous Stuart-Landau systems. The same patterns of partial amplitude death and partial synchronization have been observed for heterogeneous frequencies.
This broad appearance of the dynamical patterns predicted by the eigensolutions raises the question if there is a general concept present that one may be able to use in more complex systems.

Finally, the stability and occurrence has been fully linked to the network topology for complete in-phase synchronization as well as for patterns of interchanging in-phase and anti-phase synchronized oscillators on bipartite graphs. Their stability is shown to depend solely on two of the network's topological eigenvalues while their existence is independent of the network's size and shape. This implies that the topology can be used to (de)stabilize these states, an aspect linked to the control on complex systems that should be further investigated.
\section{Acknowledgments}
This work was supported by DFG in the framework of SFB 910. Helpful discussions with Wolfram Just, Kenneth Showalter and Mark Tinsley are gratefully acknowledged.
\appendix
\section{Eigenvalues of the Coupling Matrix}  \label{AppendixA}
The adjacency matrix describes the topology of a network. It is usually defined as 
\begin{equation}
 \tilde{A}_{ij}=\left\{\begin{array}{l}
                   1 \text{ if node }i\text{ and }j\text{ are connected}\\
                   0 \text{ otherwise.}
                  \end{array}\right.
\label{eq:tildea}
\end{equation}
where $i,j\in\{1,\dots,N\}$.
For the purpose of this study we use a definition where $A$ is normalized to unity row sum
\begin{equation}
A=D^{-1}\tilde{A}~,
\label{eq:a}
\end{equation}
with the degree matrix $D$
\begin{equation*}
D=\text{diag}(d_1,d_2,\cdots,d_n)
\end{equation*}
where $d_i$ is the degree of node $i$. 
 $A$ is in general not symmetric, but $D^{-1}$ and (for undirected networks) $\tilde{A}$ are. Note that $\tilde{A}$ is symmetric with only real entries and that $D$ is obviously symmetric, real and positive definite 
since all eigenvalues $\frac{1}{d_i}>0$. It therefore can be written in Cholesky decomposition as
\begin{equation*}
 D=LL^T
\end{equation*}
where $L$ is real and $L^T$ is the transpose of $L$. Now we write
\begin{align*}
 A&=D^{-1}\tilde{A}\\
 &=LL^T\tilde{A}
\end{align*}
and introduce the matrix
\begin{equation}B=L^T\tilde{A}L~.\label{eq:B}\end{equation}
It is similar to $A$, i.e., there exists a matrix $P$ such that 
\begin{align}B=P^{-1}AP~.\label{eq:BP}\end{align}
It is easy to see that $P=L$ satisfies this condition
\begin{align*}P^{-1}AP=L^{-1}\left(LL^T\tilde{A}\right)L=L^T\tilde{A}L=B~.\end{align*} 
Now we show that two similar matrices have the same eigenvalues. If $\nu$ is an eigenvector of $A$ with eigenvalue $\lambda$
\begin{equation*}A\nu=\lambda\nu\end{equation*}
then using $A=PBP^{-1}$ obtained from Eq. (\ref{eq:BP}) we can write
\begin{align*}PBP^{-1}\nu&=\lambda\nu\\
\Rightarrow BP^{-1}\nu&=\lambda P^{-1}\nu~.\end{align*}
Thus, if $\nu$ is an eigenvector of $A$ with eigenvalue $\lambda$, then $P^{-1}\nu$ is an eigenvector of $B$ with the same eigenvalue 
$\lambda$. Similarly, if
\begin{equation*}B\nu=\lambda\nu\end{equation*}
we can use $B=P^{-1}AP$ and obtain
\begin{align*}P^{-1}AP\nu&=\lambda\nu\\
\Rightarrow AP\nu&=\lambda P\nu~.\end{align*}
 Hence it is true that $A$ and $B$ have the same eigenvalues. 
Now note that according to Eq. (\ref{eq:B}) $B$ is real and as can be seen from
\begin{equation*}
 B^T=\left(L^T\tilde{A}L\right)^T=L^TA^T(L^T)^T=L^TAL=B
\end{equation*}
it is also symmetric. This means $B$ and therefore the normalized adjacency matrix $A$ has only real eigenvalues. 

In addition, the Gershgorin circle theorem yields that since $\sum_jA_{ij}=1~\forall~i\in\{1,\dots,N\}$ the eigenvalues of $A$ lie inside the unit circle in the complex plane. In summary:
\begin{equation}
-1\leq\eta\leq1.
\end{equation}

\section{Eigenvalues of the matrix $VAV$}\label{app:B}
Let $\eta$ be an eigenvalue of $A$,
\begin{equation*}
 A\mathbf{u}=\eta\mathbf{u}~.
\end{equation*}
Further, it follows from $v_i=\pm1$ that $V=V^{-1}$ and therefore
\begin{align*}
 VA(V^{-1}V)\mathbf{u}&=V\eta\mathbf{u}\\
 \Rightarrow VAV(V\mathbf{u})&=\eta(V\mathbf{u})~.
\end{align*}
Thus, we found an eigenvector, $\mathbf{w}=V\mathbf{u}$, of $VAV$ with eigenvalue $\eta$ and therefore each eigenvalue of $A$ is an eigenvalue of $VAV$. The eigenvectors $\mathbf{u}^i$ of $A$ are linearly independent, and since $V$ is invertible the set of eigenvectors $\mathbf{w^i}$ are as well linearly independent. 
\bibliography{ref} {}

\end{document}